\documentclass[10pt]{article}

\usepackage{amsmath}
\usepackage{graphicx}
\usepackage{orcidlink}
\usepackage{hyperref}
\usepackage{subcaption}
\usepackage{float}
\usepackage{multirow}
\usepackage{colortbl}

\RequirePackage[numbers,sort&compress]{natbib}
\begin{document}

\baselineskip=16pt

\begin{center}
{\Large \textbf{Geodesics and scalar perturbations of Schwarzschild black
holes embedded in a Dehnen-type dark matter halo with quintessence}}
\end{center}

\vspace{0.3cm}
\begin{center}
\textbf{B. Hamil}\orcidlink{0000-0002-7043-6104}\\[0pt]
Laboratoire de Physique Math\'{e}matique et Subatomique,\\
Facult\'{e} des Sciences Exactes, Universit\'{e} Constantine 1, Constantine, Algeria.\\[0pt]
e-mail: hamilbilel@gmail.com\\[0pt]
\end{center}
\begin{center}
\textbf{Ahmad Al-Badawi}\orcidlink{0000-0002-3127-3453}\\[0pt]
Department of Physics, Al-Hussein Bin Talal University, 71111, Ma'an,
Jordan. \\[0pt]
e-mail: ahmadbadawi@ahu.edu.jo\\[0pt]
\end{center}

\begin{center}
\textbf{B. C. L\"{u}tf\"{u}o\u{g}lu}\orcidlink{0000-0001-6467-5005}\\[0pt]
Department of Physics, Faculty of Science, University of Hradec Králové,\\
Rokitanskeho 62/26, Hradec Králové, 500 03, Czech Republic.\\[0pt]
e-mail: bekir.lutfuoglu@uhk.cz \\[0pt]
\end{center}

\vspace{0.3cm}
\begin{abstract}

We perform a thorough analysis into a Schwarzschild black hole
embedded in a Dehnen-type dark matter halo  with a quintessential field. We develop the composite
spacetime metric and examine its geometric properties, including horizon
structure and curvature invariants. Our findings reveal that increasing both
the DM core density $\rho _{s}$ and quintessence parameter $c$
leads to an expansion of the event horizon and a reduction in the size of
the cosmological horizon. We then investigate the dynamics of timelike and
null geodesics, focusing on the determination of innermost stable circular
orbits, photon sphere radii, and black hole shadow features. Thereafter,
using the Gauss-Bonnet theorem, we calculate the weak deflection angles,
demonstrating that lensing effects are enhanced with increasing halo density
and radius. Scalar perturbations are examined using the sixth-order WKB
method and Pad\'{e} approximants, highlighting suppressed quasinormal mode
frequencies as DM density rises. Greybody factors and Hawking
radiation sparsity are also explored, showing increased transmission
coefficients for larger halos and deviations from standard blackbody
behavior. These results underscore the significant influence of DM
and quintessence on black hole observables, offering testable predictions
for astrophysical probes such as Event Horizon Telescope imaging and
gravitational wave spectroscopy. Scalar perturbations are analyzed using the
6th-order WKB method, demonstrating that quasinormal mode frequencies are
suppressed as the DM density increases. We also explore greybody
factors and the sparsity of Hawking radiation, showing increased
transmission coefficients for larger halos and deviations from standard
blackbody behavior.
\end{abstract}

\section{Introduction}

Black holes (BHs) remain among the most fascinating and theoretically rich predictions of general relativity. Originally formulated as vacuum solutions to Einstein's field equations, they are now understood to inhabit complex astrophysical environments, often interacting with various forms of surrounding matter and fields. In particular, there is growing interest in understanding the interplay between BHs and the two dominant but mysterious components of the cosmos: dark matter (DM) and dark energy (DE). Breakthrough observations—such as the imaging of BH shadows in M87* and Sgr A* by the Event Horizon Telescope (EHT)~\cite{eht2019,eht2022} and gravitational wave detections by LIGO/Virgo~\cite{ligo2016}—have underscored the need for more realistic BH models that incorporate these dark sector fields.

Theoretical and observational studies consistently support the existence of extended DM halos surrounding galaxies, inferred from galactic rotation curves, strong and weak lensing, and cosmic microwave background anisotropies. These halos are thought to comprise approximately 27\% of the total energy density of the universe. One widely used model for describing such halos is the Navarro–Frenk–White (NFW) profile~\cite{nfw1996}, which arises naturally from cold DM (CDM) simulations and predicts a characteristic cuspy inner density slope. However, the NFW model lacks flexibility in accommodating alternative halo structures such as cored distributions. In contrast, the Dehnen profile~\cite{dehnen1993} is a more versatile double power-law density distribution that includes the Jaffe and Hernquist profiles as special cases and allows for tunable central slope parameters. This flexibility makes the Dehnen profile particularly well-suited for modeling DM environments around BHs in both cuspy and cored configurations~\cite{macmillan1999,stegmann2020}.

Recent theoretical efforts have focused on embedding BHs within Dehnen-type DM backgrounds. The authors in ~\cite{badawi2024} developed a Schwarzschild-like solution surrounded by a Dehnen-(1,4,5/2) halo and analyzed its geometric and causal structure. They found that while the solution preserves asymptotic flatness and central singularity, the DM halo induces notable deviations in effective potentials and event horizons. Extending this line of inquiry, Gohain et al.~\cite{gohain} studied the thermodynamics and null geodesics of a Schwarzschild-Dehnen system, demonstrating that parameters like core radius and DM density substantially impact Hawking temperature, photon spheres, and the specific heat capacity. 

Parallel investigations have also explored the role of DE in BH physics. One widely used approach is the Kiselev model~\cite{kiselev2003}, which incorporates quintessence via a fluid with an equation of state $p_q = \omega_q \rho_q$, where $-1 < \omega_q < -1/3$. This phenomenological model captures the repulsive nature of DE and allows its localized influence on BH solutions to be studied analytically. Comprehensive reviews of DE models and their gravitational implications can be found in~\cite{nojiri2011,nojiri2017}. 
Additional phenomenological insights arise from the analysis of test particle motion and quasinormal modes (QNMs). Xamidov et al.~\cite{xamidov} examined quasi-periodic oscillations (QPOs) and innermost stable circular orbits (ISCOs) around Schwarzschild BHs in Dehnen halos, comparing their results with X-ray observations from microquasars. In another study, ~\cite{badawi2025} focused on scalar perturbations and timelike geodesics in Schwarzschild–Dehnen–quintessence spacetimes, calculating QNMs and identifying observable deviations in geodesic structure.

Motivated by these developments, the present work provides a unified and comprehensive study of a Schwarzschild-like BH immersed in both a Dehnen-type DM halo and a surrounding quintessence field. We derive the composite spacetime metric and analyze its geometric features, including horizon structure, curvature invariants, and the classical energy conditions. We then investigate geodesic dynamics, focusing on effective potentials, circular orbits, and ISCOs. Additionally, we examine the BH shadow and weak gravitational lensing using the Gauss-Bonnet theorem, and explore scalar field perturbations and QNMs via the sixth-order WKB method. Together, these analyses provide insight into the imprints of dark sector fields on BH observables and offer predictions that may be testable with current and upcoming astrophysical instruments.

The remainder of this paper is organized as follows. In Section~\ref{sec:geometry1}, we present the theoretical framework and derive the spacetime metric for a Schwarzschild-like BH embedded in a Dehnen-type DM halo with a surrounding quintessence field. Further, we analysis of the spacetime geometry, including horizon structure, curvature invariants, and the satisfaction of classical energy conditions. In Section~\ref{sec:geodesics}, we study the geodesic motion of test particles and light, focusing on effective potentials, circular orbits, and the innermost stable circular orbit (ISCO). Section~\ref{sec:shadow} explores the optical properties of the BH, including shadow formation and gravitational lensing in the weak field limit. In Section~\ref{sec:qnm}, we investigate scalar field perturbations and compute the quasinormal modes using the sixth-order WKB method. Finally, in Section~\ref{sec:conclusion}, we summarize our findings and discuss their physical implications and potential observational relevance.


\section{Spacetime metric} \label{sec:geometry1}

\begin{figure}[H]
    \centering
    \includegraphics[width=0.5\linewidth]{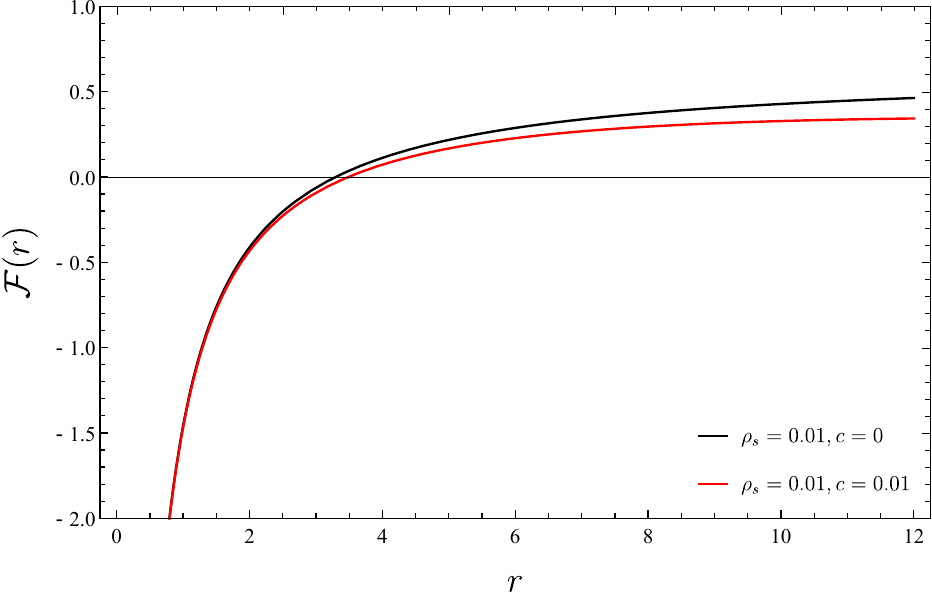}
    \caption{ Lapse function $\mathcal{F}(r)$ as a function of $r$. It shows that the BH has two horizons. Here, we use $M=1$, $r_s=0.6$ and $w_q=-2/3$.}
    \label{lapse1}
\end{figure}

\begin{figure}[H]
    \centering
    \includegraphics[width=0.45\linewidth]{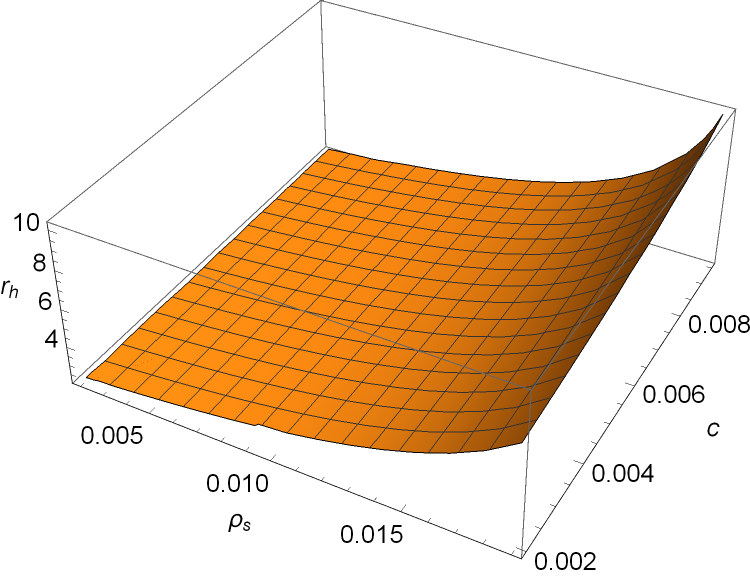}\quad\quad
    \includegraphics[width=0.45\linewidth]{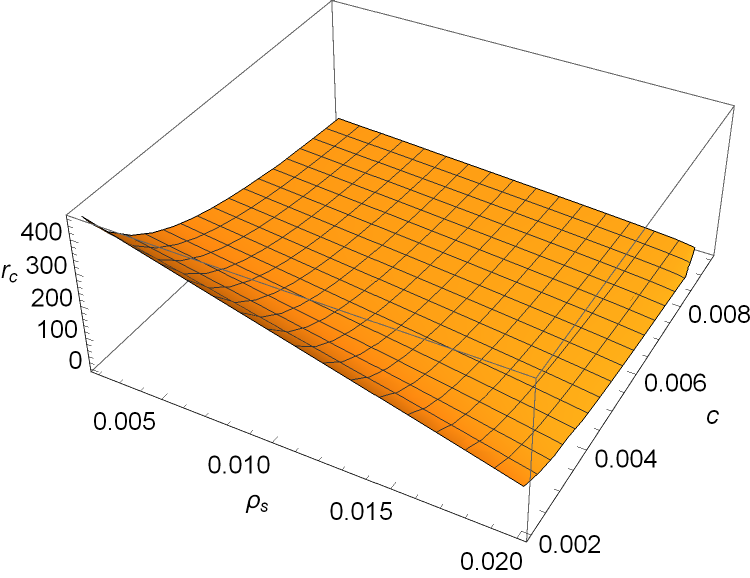}
    \caption{ Variation of the event (left) and cosmological (right) horizons  with the parameters $\rho$ and $c$. Here, we use $M=1$, $r_s=0.6$ and $w_q=-2/3$.}
    \label{Horiz1}
\end{figure}

In Ref. \cite{badawi2024}, a Schwarzschild-like BH (BH) solution
describing a static and asymptotically flat BH surrounded by a DM
(DM) halo with a Dehnen-type density distribution in the surrounding
environment was reported. Building upon this work, the present study extends
the analysis by incorporating the quintessence field (QF) into the BH
solution. This extension involves a comprehensive investigation of the
geodesic motion of test particles, scalar perturbations, and the
thermodynamics of the system. Specifically, we introduce a static and
spherically symmetric Schwarzschild-like BH in presence of QF, which is
described by the following line element: 
\begin{equation}
ds^{2}=-\mathcal{F}(r)\,dt^{2}+\frac{dr^{2}}{\mathcal{F}(r)}+r^{2}\,(d\theta
^{2}+\sin ^{2}\theta \,d\varphi ^{2}),  \label{bb1}
\end{equation}%
where the metric function $\mathcal{F}(r)$ is 
\begin{equation}
\mathcal{F}(r)=1-\frac{2\,M}{r}-32\pi \rho _{s}r_{s}^{3}\sqrt{\frac{r+r_{s}}{%
r_{s}^{2}\,r}}-\frac{c}{r^{3\,\omega _{q}+1}}  \label{bb2}
\end{equation}%
 where  $\rho_s$ and $r
 _s$ are the central halo density radius, ($%
c,\omega _{q}$) are QF parameters and $\rho _{s}$ is the central halo
density, $r_{s}$ is the halo core radius. In the limit where  $c=0$, the
metric (\ref{bb1}) reduces to a Schwarzschild-like BH \cite{badawi2024}.\\ To better understand the metric function (\ref{bb2}), we plot it as a function of $r$, as shown in Figure \ref{lapse1}. The graph shows that there are only two horizons for BH with a Dehnen type DM halo and QF, namely the event ($r_h$) and cosmological ($r_c$) horizons. The BH has a distinct horizon when $c=0$, i.e., no QF. 
The horizons can
 be determined using the condition  \begin{equation}
    1-\frac{2\,M}{r}-32\pi \rho _{s}r_{s}^{3}\sqrt{\frac{r+r_{s}}{%
r_{s}^{2}\,r}}-\frac{c}{r^{3\,\omega _{q}+1}} =0. \label{hor1} 
 \end{equation}
This equation (\ref{hor1}) has no analytical solution.  We use numerical methods to determine the event and cosmological horizons. In Figure \ref{Horiz1}, we depict both horizons and demonstrate how parameters core density ($\rho$) and QF ($c$) influence them. Figure \ref{Horiz1} indicates that $\rho$ and $c$ have similar effects, with the event horizon increasing with both (see left panel). However, there is an effect on the cosmological horizon. The cosmological horizon shrinks substantially when $\rho$ and $c$ rise (see right panel).
Curvature scalars can be used to explore the behavior of the BH solution. Kretschmann's scalar for metric (\ref{bb1}) is
\begin{eqnarray}
    \mathcal{K}&=& \frac{4}{r^6}\left(2M+cr^{-3w_q}(1+3w_q) \frac{16\pi r_s^3\rho_s}{\sqrt{\frac{r+r_{s}}{%
r}}}  \right)^2+r^{-6w_q}\left( c+2r^{3w_q} \left( M+16\pi r_s^2 \rho \sqrt{\frac{r+r_{s}}{%
r}} \right)^2 \right)  \nonumber \\  &+&\frac{1}{r^8}\left( 4Mrcr^{1-3w_q}(2+9w_q(1+w_q))+\frac{8\pi r_s^3 \rho(4r+3r_s) }{\left( \frac{r+r_{s}}{
r} \right)^{3/2} }      \right)^2.
\end{eqnarray}
The Kretschmann scalar has a singularity at $r=0$. 
 Thus, the BH solution's singularity at $ r=0$ is an inherent singularity that no coordinate modification can remove.

\bigskip

\section{Geodesics Equation} \label{sec:geodesics}

In this section, we will derive the equations of motion and explore the characteristics of null \ and time-like geodesics for Schwarzschild BHs situated within a Dehnen-type DM halo with quintessence. We firstly set up the geodesic equations and their corresponding constraint conditions which are given by
\begin{equation}
\ddot{x}^{\mu }+\Gamma _{\alpha \beta }^{\mu }\dot{x}^{\alpha }\dot{x}%
^{\beta }=0,
\end{equation}%
\begin{equation}
g_{\alpha \beta }\dot{x}^{\alpha }\dot{x}^{\beta }=-\epsilon .
\end{equation}%
In these equations, $\dot{x}$ represent the differentiation with respect to the affine parameter $\lambda ,$ $\epsilon =0$ and $1$ for null and time-like geodesics. The geodesic and constraint on the trajectories equations for the spherically symmetric spacetimes take the following forms%

\begin{equation}
\ddot{t}+\frac{\mathcal{F}^{\prime }(r)}{\mathcal{F}(r)}\dot{t}\dot{r}=0,
\end{equation}%
\begin{equation}
\ddot{r}+\frac{1}{2}\mathcal{F}(r)\mathcal{F}^{\prime }(r)\dot{t}^{2}-\frac{%
\mathcal{F}^{\prime }(r)}{2\mathcal{F}(r)}\dot{r}^{2}-r\mathcal{F}(r)\left( 
\dot{\theta}^{2}+\dot{\varphi}^{2}\sin ^{2}\theta \right) =0,
\end{equation}%
\begin{equation}
\ddot{\theta}+\frac{2}{r}\dot{r}\dot{\theta}-\sin \theta \cos \theta \dot{%
\varphi}^{2}=0,
\end{equation}%
\begin{equation}
\ddot{\varphi}+\frac{2}{r}\dot{r}\dot{\varphi}+2\dot{\theta}\dot{\varphi}%
\cot \theta =0,
\end{equation}%
\begin{equation}
-\mathcal{F}(r)\,\dot{t}^{2}+\frac{\dot{r}^{2}}{\mathcal{F}(r)}+r^{2}\,(\dot{%
\theta}^{2}+\sin ^{2}\theta \,\dot{\varphi}^{2})=-\epsilon .
\end{equation}%
Now, we focus on the equatorial plane ($\theta =\pi /2$). Under this
condition, the geodesic equations and their corresponding constraint
equation can be simplified and rewritten as follows:%
\begin{equation}
\ddot{t}+\frac{\mathcal{F}^{\prime }(r)}{\mathcal{F}(r)}\dot{t}\dot{r}=0,
\label{1}
\end{equation}%
\begin{equation}
\ddot{r}+\frac{1}{2}\mathcal{F}(r)\mathcal{F}^{\prime }(r)\dot{t}^{2}-\frac{%
\mathcal{F}^{\prime }(r)}{2\mathcal{F}(r)}\dot{r}^{2}-r\mathcal{F}(r)\dot{%
\varphi}^{2}=0,
\end{equation}%
\begin{equation}
\ddot{\varphi}+\frac{2}{r}\dot{r}\dot{\varphi}=0,  \label{3}
\end{equation}%
\begin{equation}
-\mathcal{F}(r)\,\dot{t}^{2}+\frac{\dot{r}^{2}}{\mathcal{F}(r)}+r^{2}\,\dot{%
\varphi}^{2}=-\epsilon .  \label{cons}
\end{equation}%
By integrating Eqs. (\ref{1}, \ref{3}) leads to%
\begin{equation}
\dot{t}=\frac{E}{\mathcal{F}(r)},  \label{4}
\end{equation}%
\begin{equation}
\dot{\varphi}=\frac{L}{r^{2}}.  \label{5}
\end{equation}%
Here $E$ and $L$ are nd the energy the angular momentum respectively.
Simplifying the Eq. (\ref{cons}), by substituting Eqs. (\ref{4}, \ref{5})
into Eq. (\ref{cons}), we can obtain

\begin{equation}
\dot{r}^{2}\,=E^{2}-\left( 1-\frac{2\,M}{r}-32\pi \rho _{s}r_{s}^{3}\sqrt{%
\frac{r+r_{s}}{r_{s}^{2}\,r}}-\frac{c}{r^{3\,\omega _{q}+1}}\right) \left(\epsilon +\frac{L^{2}}{r^{2}}\right) .
\end{equation}%

This allows us to rewrite the above equation into a one-dimensional form%
\begin{equation}
\dot{r}^{2}\,=E^{2}-V_{\mathrm{eff}}\left( r\right) ,
\end{equation}

where the effective potential is defined as%
\begin{equation}
V_{\mathrm{eff}}\left( r\right) =\left( 1-\frac{2\,M}{r}-32\pi \rho
_{s}r_{s}^{3}\sqrt{\frac{r+r_{s}}{r_{s}^{2}\,r}}-\frac{c}{r^{3\,\omega
_{q}+1}}\right) \left( \epsilon +\frac{L^{2}}{r^{2}}\right) , \label{potentiel}
\end{equation}%
which governs the radial motion of the particle in the equatorial plane.

\subsection{Time-like geodesics motion}
\begin{figure}[H]
\begin{minipage}[t]{0.33\textwidth}
        \centering
        \includegraphics[width=\textwidth]{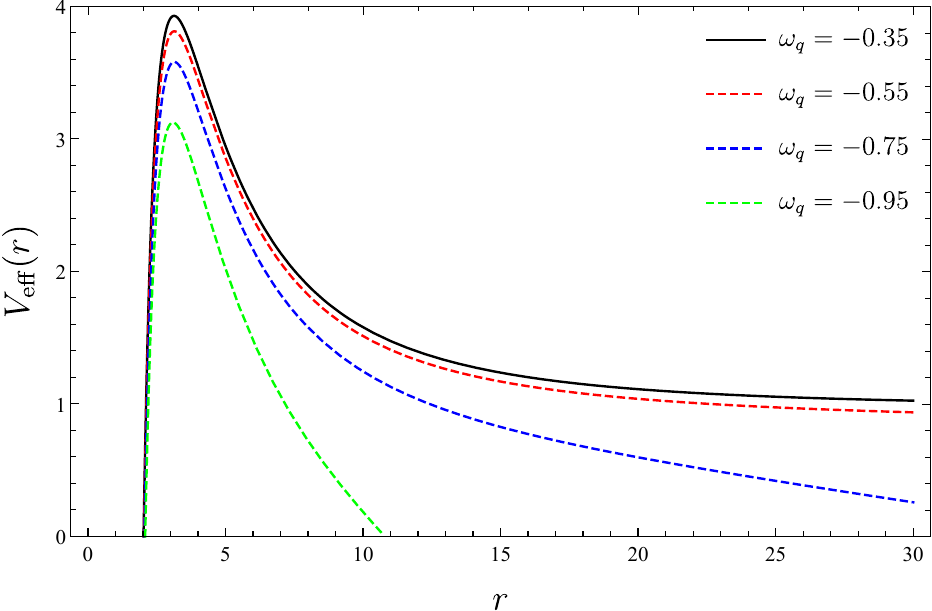}
       \subcaption{ $ \rho_{s}=0$.}\label{fig:veff1}
   \end{minipage}%
\begin{minipage}[t]{0.33\textwidth}
       \centering
        \includegraphics[width=\textwidth]{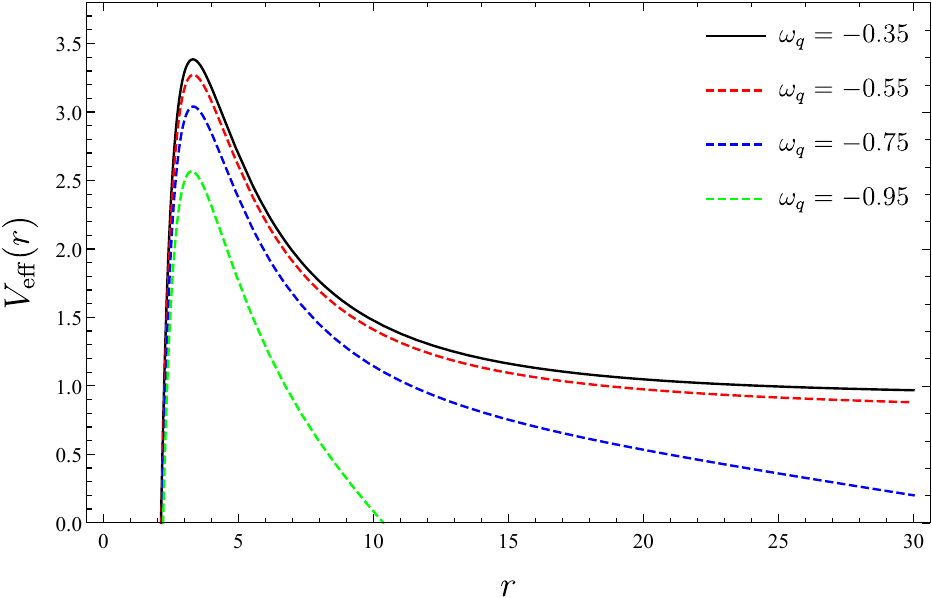}
       \subcaption{ $ \rho_{s}=0.05$}\label{fig:veff2}
   \end{minipage}
   \begin{minipage}[t]{0.33\textwidth}
        \centering
        \includegraphics[width=\textwidth]{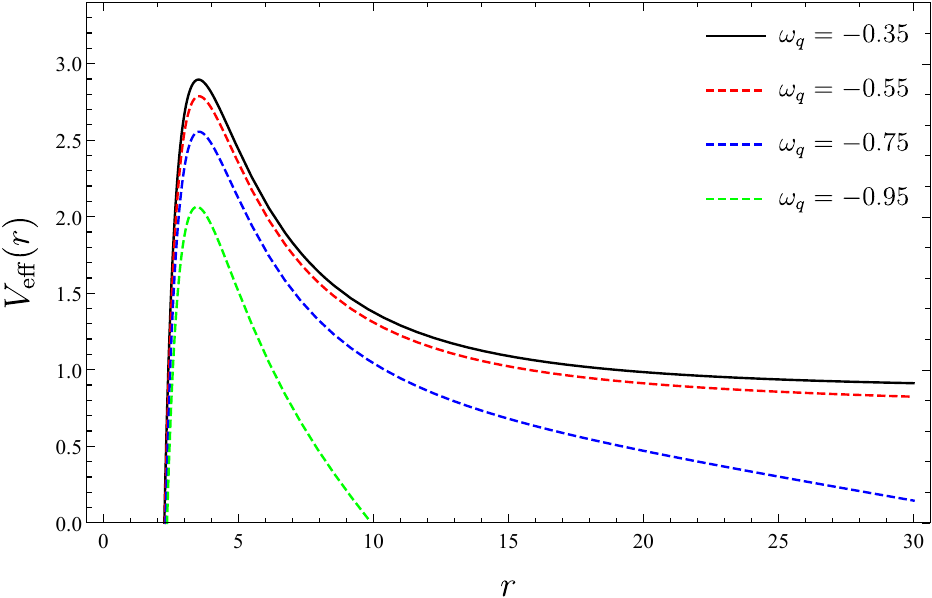}
         \subcaption{ $ \rho_{s}=0.1$.}\label{fig:veff3}
   \end{minipage}
\caption{The behaviors of timelike geodesic effective potential  $V_{\mathrm{eff}}\left( r\right) $ as a function of 
$r$ for $L=10,M=1,c=0.01$ and $r_{s}=0.1.$}
\label{figveff1}
\end{figure}

\begin{figure}[H]
    \centering
    \includegraphics[width=0.7\linewidth]{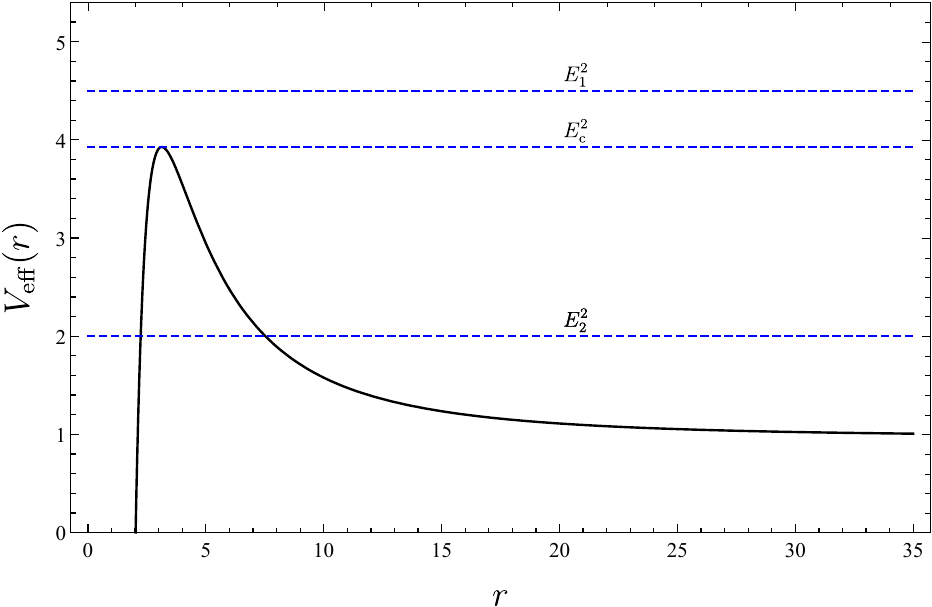}
    \caption{The behavior of $V_{\mathrm{eff}}\left( r\right) $ as a function of 
$r$ for $L=10,M=1,c=0.01$ and $\rho_{s}=0.$}
    \label{fig:typesoforbits}
\end{figure}

\begin{figure}[H]
\begin{minipage}[t]{0.5\textwidth}
        \centering
        \includegraphics[width=\textwidth]{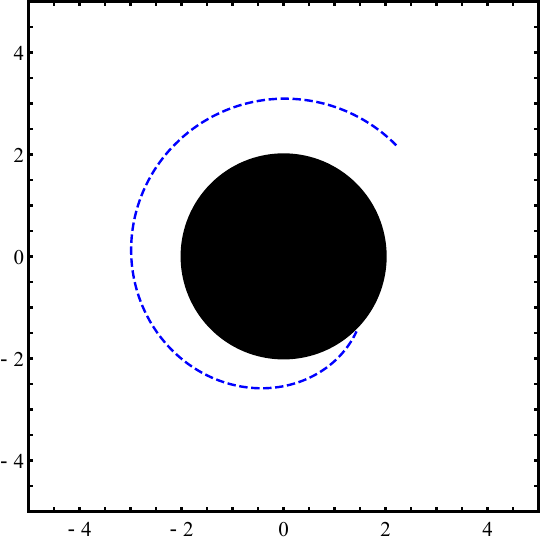}
       \subcaption{ $\rho_{s}=0$.}\label{fig:tr1}
   \end{minipage}%
\begin{minipage}[t]{0.5\textwidth}
       \centering
        \includegraphics[width=\textwidth]{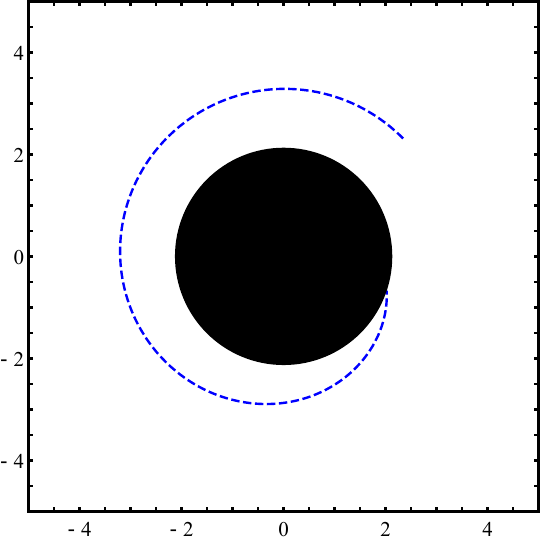}
       \subcaption{ $\rho_{s}=0.05$}\label{fig:tr2}
   \end{minipage}
\caption{Plots of the timelike unstable circular orbits }
\label{figtrajectoire1}
\end{figure}

The effective potential is crucial for analyzing the trajectories of test particles, allowing us to describe their motions without directly solving the equations of motion. The extrema of the effective potential distinguish between stable and unstable circular orbits: maxima correspond to unstable orbits, while minima indicate stable ones. For timelike trajectories, the effective potential can be written as:
\begin{equation}
V_{\mathrm{eff}}\left( r\right) =\left( 1-\frac{2\,M}{r}-32\pi \rho
_{s}r_{s}^{3}\sqrt{\frac{r+r_{s}}{r_{s}^{2}\,r}}-\frac{c}{r^{3\omega_{q}+1}}\right) \left( 1+\frac{L^{2}}{r^{2}}\right) ,\label{effp1}
\end{equation}
It is important to note that the effective potential is influenced by several parameters, such as the mass $M$, angular momentum $L$, central halo density $\rho_{s}$, halo core radius parameter $r_{s}$, and the QF parameters ($c,\omega_{q}$). By setting the parameters $M=1$, $L=10$, $r_{s}=0.1$ and $c=0.01$ we show the effective potential curves for various values of $\rho_{s}$ and $\omega_{q}$ in Fig. \ref{figveff1}. Our analysis reveals that the Dehnen-type DM halo and the quintessence field significantly influence the effective potential. As the parameter $\omega_{q} $ increases, the effective potential energy shifts to the right, and its maximum value also rises. Additionally, we observe that the maximum value of the effective potential energy decreases with higher values of the central halo density $\rho_{s}$. 
We now examine the trajectories of the radial particle in detail, focusing on various types of orbits corresponding to different energy values $E$, as illustrated in Fig. \ref{fig:typesoforbits}:
\begin{itemize}
\item When $E^{2} > E^{2}_{c}$, the particle starting from rest, will fall directly into the center. 
\item  When $E^{2} = E^{2}_{c}$,  the particle follows an unstable circular orbit and may eventually spiral into the singularity, depending on its initial energy conditions. This kind of unstable circular orbit is illustrated in Fig. \ref{figtrajectoire1}
\item When $E^{2}_{2}<E^{2} < E^{2}_{c}$, the particle comes from infinity, moves toward the center, and after reaching a closest point, it then returns to infinity.
\end{itemize}
We then examine the behavior of timelike particles in circular orbits around a BH embedded in a Dehnen-type DM halo with quintessence. In order for the particles to remain on these circular orbits, the following conditions must be fulfilled simultaneously
\begin{equation}
V_{\mathrm{eff}}\left( r\right) =E^2 \quad \text{and} \quad   \partial_{r} V_{\mathrm{eff}}\left( r\right) =0.
\end{equation}
The equation above can be solved to determine the radii of circular orbits for specific values of energy $E$ and angular momentum $L$. The radial distributions of the angular momentum $L$ and energy $E$ for these circular orbits are expressed in the following form:
\begin{equation}
L^{2}=\frac{r}{2}\frac{2m+\frac{c(3\omega _{q}+1)}{r^{3\omega _{q}}}+16\pi\frac{\rho _{s}r_{s}^{2}}{r}\sqrt{\frac{r_{s}^{2}\,r}{r+r_{s}}}}{1-\frac{3m}{r}-32\pi \rho_{s}r_{s}\sqrt{\frac{r_{s}^{2}\,r}{r+r_{s}}}-40\pi \frac{\rho
_{s}r_{s}^{2}}{r}\sqrt{\frac{r_{s}^{2}\,r}{r+r_{s}}}-\frac{3}{2}\frac{c(\omega _{q}+1)}{r^{1+3\omega _{q}}}}.\label{554}
\end{equation}

\begin{equation}
E^{2}=\frac{\left( 1-\frac{2M}{r}-\frac{c}{r^{1+3\omega _{q}}}-32\pi \rho
_{s}r_{s}^{3}\sqrt{\frac{r+r_{s}}{r_{s}^{2}\,r}}\right) ^{2}}{1-\frac{3M}{r}-32\pi \rho _{s}r_{s}\sqrt{\frac{r_{s}^{2}\,r}{r+r_{s}}}-40\pi \frac{\rho
_{s}r_{s}^{2}}{r}\sqrt{\frac{r_{s}^{2}\,r}{r+r_{s}}}-\frac{3c(\omega _{q}+1)}{2r^{1+3\omega _{q}}}\sqrt{\frac{r_{s}^{3}\,r}{r+r_{s}}}}.\label{51}
\end{equation}

In astrophysics, the study of test particle motion around compact objects is driven by the investigation of a particular type of circular orbit known as the innermost stable circular orbit (ISCO). This orbit is crucial for understanding massive compact objects like neutron stars, quasars, and supermassive BHs. The radius of the ISCO can be determined using the following condition:
\begin{equation}
\frac{\partial^2 V_{\mathrm{eff}}\left( r\right)}{\partial r^2}=0. \label{53}  
\end{equation}
By solving Eq. (\ref{53}) together with Eqs. (\ref{51}) and (\ref{554}), the following condition is obtained:
\begin{eqnarray}
\Biggr[ &&\,\frac{-2}{\,r}\left( 2M+\frac{c(3\omega _{q}+1)}{r^{3\omega _{q}}%
}+16\pi \rho _{s}r_{s}^{2}\sqrt{\frac{r_{s}^{2}\,r}{r+r_{s}}}\right) ^{2} 
\notag \\
&&-\left( 1-\frac{c}{r^{1+3\omega _{q}}}-\frac{2M}{r}-32\pi \rho r_{s}^{3}%
\sqrt{\frac{r+r_{s}}{r_{s}^{2}\,r}}\right) \times \left( \frac{c(3\omega
_{q}+1)(3\omega _{q}+2)}{r^{3\omega _{q}}}+4M+8\pi \rho _{s}r_{s}^{2}(\frac{%
4r+3r_{s}}{r+r_{s}})\sqrt{\frac{r_{s}^{2}\,r}{r+r_{s}}}\right)   \notag \\
&&+3\left( 1-\frac{c}{r^{1+3\omega _{q}}}-\frac{2M}{r}-32\pi \rho
_{s}r_{s}^{3}\sqrt{\frac{r+r_{s}}{r_{s}^{2}\,r}}\right) \left( 2m+\frac{%
c(3\omega _{q}+1)}{r^{3\omega _{q}}}+16\pi \rho _{s}r_{s}^{2}\sqrt{\frac{%
r_{s}^{2}\,r}{r+r_{s}}}\right) \Biggr]_{r=r_{\mathrm{ISCO}}}=0.\label{72}
\end{eqnarray}

The exact solution of Eq. (\ref{72}) cannot be obtained, so we proceed with a numerical approach. The presence of the central halo density introduces an additional parameter, $\rho_{s}$, in Eq. (\ref{72}). We examine 6-values for $\rho_{s}$: 0, 0.01, 0.02, 0.03, 0.04, and 0.05. Using these values, we numerically solve Eq. (\ref{72}) to determine the ISCO radius, $r_{\mathrm{ISCO}}$. The ISCO parameters $r_{\mathrm{ISCO}}$, $L_{\mathrm{ISCO}}$ and $E_{\mathrm{ISCO}}$ are computed numerically and shown in Table \ref{tab1}.

\begin{table}[H]
\centering
\begin{tabular}{|l|lll|lll|lll|}
\hline
\rowcolor{lightgray} \multirow{2}{*}{} & \multicolumn{3}{|c|}{$\omega
_{q}=-0.35$} & \multicolumn{3}{|c|}{$\omega _{q}=-0.45$} & 
\multicolumn{3}{|c|}{$\omega _{q}=-0.55$} \\ \hline\hline
\rowcolor{lightgray} $\rho _{s}$ & $r_{\mathrm{ISCO}}$ & $L_{\mathrm{ISCO}}$ & $E_{\mathrm{ISCO}}$ & $%
r_{\mathrm{ISCO}}$ & $L_{\mathrm{ISCO}}$ & $E_{\mathrm{ISCO}}$ & $r_{\mathrm{ISCO}}$ & $L_{\mathrm{ISCO}}$ & $E_{\mathrm{ISCO}}$
\\ \hline
$0$ & 6.06673 & 3.49855 & 0.937377 & 6.14768 & 3.48312 & 0.930756 & 6.46874
& 3.42136 & 0.917063 \\ 
$0.01$ & 6.13059 & 3.53534 & 0.932595 & 6.21384 & 3.51963 & 0.925897 & 
6.54714 & 3.45603 & 0.911991 \\ 
$0.02$ & 6.19578 & 3.57289 & 0.927789 & 6.28141 & 3.55689 & 0.921013 & 
6.62762 & 3.49137 & 0.906887 \\ 
$0.03$ & 6.26233 & 3.61123 & 0.922957 & 6.35044 & 3.59494 & 0.916102 & 6.71030
& 3.52741 & 0.901750 \\ 
$0.04$ & 6.33030 & 3.65038 & 0.918100 & 6.42098 & 3.63379 & 0.911163 & 6.79526
& 3.56416 & 0.896580 \\ 
$0.05$ & 6.39972 & 3.69036 & 0.913218 & 6.49309 & 3.67346 & 0.906197 & 
6.88263 & 3.60165 & 0.891376 \\ \hline\hline
\end{tabular}
\caption{Numerical estimations of the ISCO  parameters $r_{\mathrm{ISCO}}$, $L_{\mathrm{ISCO}}$ and $E_{\mathrm{ISCO}}$ for $r_{s}=0.1$, $c=0.01$ and $M=1$}
\label{tab1}
\end{table}

\subsection{Null geodesic motion: Black hole shadow}
\begin{figure}[H]
\begin{minipage}[t]{0.330\textwidth}
        \centering
        \includegraphics[width=\textwidth]{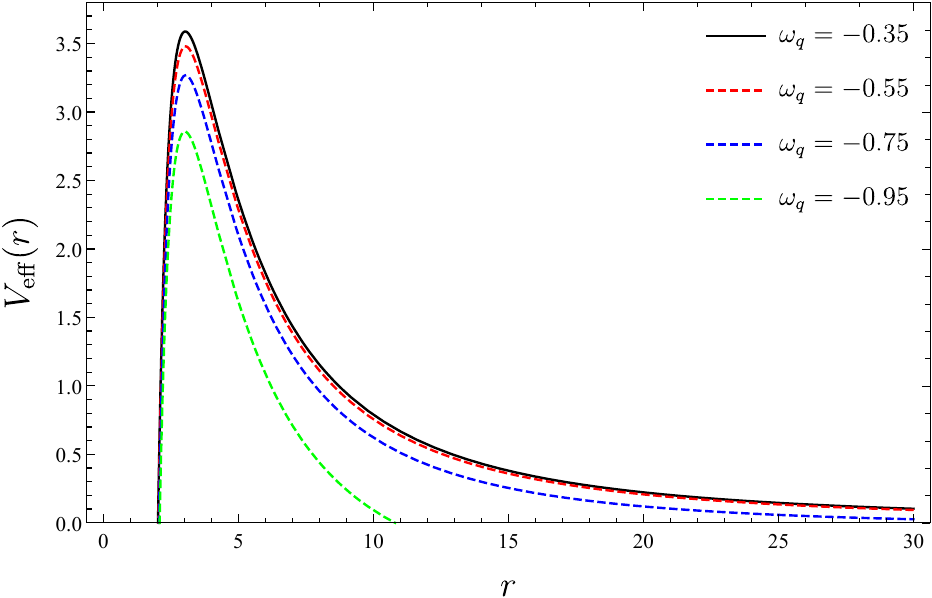}
       \subcaption{ $ \rho_{s}=0$.}\label{fig:null1}
   \end{minipage}%
\begin{minipage}[t]{0.330\textwidth}
       \centering
        \includegraphics[width=\textwidth]{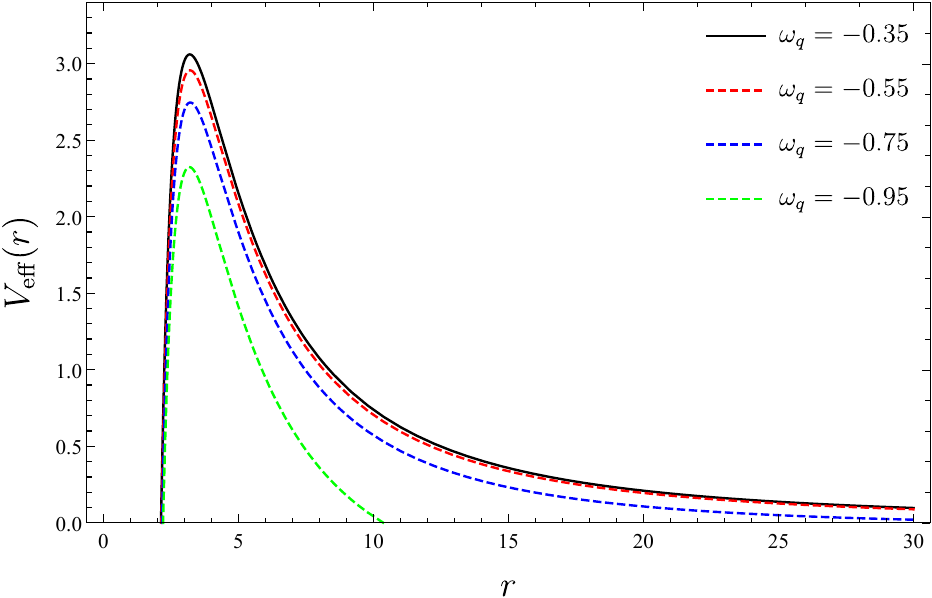}
       \subcaption{ $ \rho_{s}=0.05$}\label{fig:null2}
   \end{minipage}
\begin{minipage}[t]{0.330\textwidth}
        \centering
        \includegraphics[width=\textwidth]{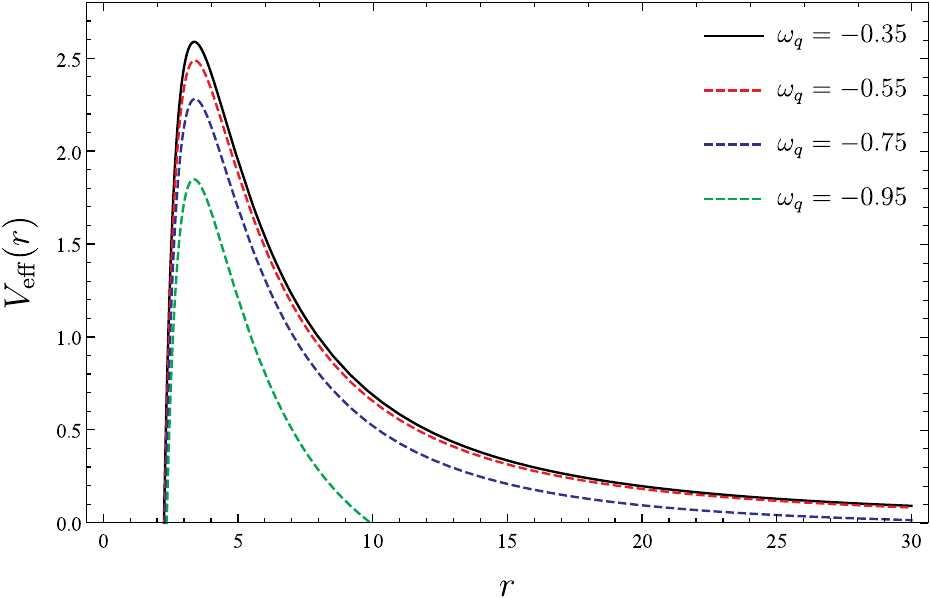}
         \subcaption{ $ \rho_{s}=0.1$.}\label{fig:null3}
   \end{minipage}
\caption{ The behaviors of null geodesic effective potential $V_{\mathrm{eff}}\left( r\right) $ as a function of 
$r$ for $L=10,M=1,c=0.01$ and $r_{s}=0.1.$}
\label{fignull}
\end{figure}

Shadows cast by BHs surrounded by various configurations of DM have been recently extensively studied (see, for instance, \cite{Konoplya:2019sns, Konoplya:2022hbl, Konoplya:2025mvj, Haroon:2018ryd, Jusufi:2019nrn} and references therein).
In this subsection, we analyze the null geodesic motion of photons around a
Schwarzschild BH surrounded by a DM halo with a Dehnen-type
density distribution and a quintessence field. The effective potential for a
photon in this spacetime is derived from (\ref{potentiel}) by setting $%
\epsilon =0$, 
\begin{equation}
V_{\mathrm{eff}}\left( r\right) =\left( 1-\frac{2\,M}{r}-32\pi \rho
_{s}r_{s}^{3}\sqrt{\frac{r+r_{s}}{r_{s}^{2}\,r}}-\frac{c}{r^{3\omega _{q}+1}}%
\right) \frac{L^{2}}{r^{2}}.
\end{equation}%
For the standard Schwarzschild case, the effective potential $V_{\mathrm{eff}%
}$ for photons reaches its maximum at $r_{\mathrm{c}}=3M$ indicating the
presence of an unstable circular orbit. As $r\rightarrow \infty $, the
effective potential asymptotically approaches a constant value. Fig.\ref%
{fignull} demonstrates the impact of the central halo density $\rho _{s}$
and the quintessence field on the effective potential of a massless test
particle (photon). One can see from Fig. \ref{fignull}, as $\rho _{s}$
increases, the photon radius $r_{\mathrm{c}}$ decreases, emphasizing that in
the presence of a DM halo with a Dehnen-type density distribution,
the unstable circular orbits shrink.

The unstable circular orbit define the boundary of the BH's apparent
shape and can be determined by finding where the effective potential reaches
its maximum value, which requires: 
\begin{equation}
V_{\mathrm{eff}}-E^{2}=\frac{\partial V_{\mathrm{eff}}}{\partial r}=0,
\end{equation}%
and check whether $V_{\mathrm{eff}}$ is a maxima at $r=r_{\mathrm{c}}$, that
is 
\begin{equation}
\frac{\partial ^{2}V_{\mathrm{eff}}}{\partial r^{2}}<0.
\end{equation}%
Using the condition $V_{\mathrm{eff}}-E^{2}=0$ leads to 
\begin{equation}
b_{\mathrm{c}}=\frac{L}{E}=\frac{r_{\mathrm{c}}}{\sqrt{1-\frac{2\,M}{r_{%
\mathrm{c}}}-32\pi \rho _{s}r_{s}^{3}\sqrt{\frac{r_{\mathrm{c}}+r_{s}}{%
r_{s}^{2}r_{\mathrm{c}}}}-\frac{c}{r_{\mathrm{c}}^{3\omega _{q}+1}}}},
\end{equation}%
while the boundary condition $\frac{\partial V_{\mathrm{eff}}}{\partial r}=0$
leads to 
\begin{equation}
1-\frac{3c+3c\omega _{q}}{2r^{1+3\omega _{q}}}-\frac{3m}{r}-\frac{8\pi \rho_{s}r_{s}^{4}}{(r+r_{s})}\sqrt{\frac{r+r_{s}}{rr_{s}^{2}}}-32\pi \rho_{s}r_{s}^{3}\sqrt{\frac{r+r_{s}}{rr_{s}^{2}}}=0.\label{30}
\end{equation}

We numerically solve Eq. (\ref{30}), and the results are presented in Table \ref{tab2}. This table shows the variation of $r_{c}$ and $b_{c}$  with respect to the quintessence parameters $\omega_{q}$ and $\rho_{s}$. From this table, we observe that for fixed values of $\rho_{s}$, the photon radius $r_{c}$ increases as the quintessence parameter $\omega_{q}$ decreases. 
\begin{table}[H]
\centering
\begin{tabular}{|l|ll|ll|ll|}
\hline
\rowcolor{lightgray} \multirow{2}{*}{} & \multicolumn{2}{|c|}{$\omega
_{q}=-0.4$} & \multicolumn{2}{|c}{$\omega _{q}=-0.6$} & \multicolumn{2}{|c|}{%
$\omega _{q}=-0.8$} \\ \hline\hline
\rowcolor{lightgray} $\rho _{s}$ & $r_{c}$ & $b_{c}$ & $r_{c}$ & $b_{c}$ & $%
r_{c}$ & $b_{c}$ \\ \hline
$0$ & 3.03409    & 5.29502 & 3.04451 & 5.39285 & 3.04337 & 5.60013 \\ 
$0.01$ & 3.06610  & 5.37848 & 3.07705 & 5.48046 & 3.07617 & 5.69850 \\ 
$0.02$ & 3.09876 & 5.46412 & 3.11029 & 5.57047 & 3.10969 & 5.79994 \\ 
$0.03$ & 3.13212 & 5.55204 & 3.14424 & 5.66298 & 3.14395 & 5.90461 \\ 
$0.04$ & 3.16619 & 5.64231 & 3.17893 & 5.75809 & 3.17897 & 6.01267 \\ 
$0.05$ & 3.20099 & 5.73502 & 3.21439 & 5.85590  & 3.21470 & 6.12428 \\ 
\hline\hline
\end{tabular}
\label{tabm}
\caption{Photon radius $r_{c}$ and the impact parameter $b_{c}$ for 3-
values of quintessence parameter $\omega _{q}=-0.4,-0.6,-0.8$ with
varying values of the DM parameter $\protect\rho _{s}$, with $M=1$, $c=0.01$ and $r_{s}=0.1$}

\label{tab2}
\end{table}
To construct the shadow, we assume a stationary observer at a radial
coordinate $r_{\mathrm{o}}$ sends light rays backward in time. The angle $%
\alpha $ between a light ray and the radial direction is given by%
\begin{equation}
\cot \alpha =\left. \frac{1}{r\sqrt{\mathcal{F}(r)}}\frac{d\varphi }{dr}%
\right\vert _{r=r_{\mathrm{o}}}.
\end{equation}

Here $r_{\mathrm{o}}$  represents the observer's distance from the black
hole. Using the geodesic equations, the above expression can be rewritten as%
\begin{equation}
\sin ^{2}\alpha =\frac{\mathcal{F}(r_{\mathrm{o}})R^{2}}{r_{\mathrm{o}}^{2}%
\mathcal{F}(R)}.  \label{22}
\end{equation}
Therefore, we must consider the limit $R\rightarrow r_{c}$  in eq. (\ref{22}%
)  to obtain the angular radius $\alpha _{\mathrm{sh}}$ of the shadow,

\begin{equation}
\sin ^{2}\alpha _{\mathrm{sh}}=\frac{r_{\mathrm{c}}^{2}\mathcal{F}(r_{%
\mathrm{o}})}{\mathcal{F}(r_{\mathrm{c}})r_{\mathrm{o}}^{2}}=b_{\mathrm{c}%
}^{2}\frac{\mathcal{F}(r_{\mathrm{o}})}{r_{\mathrm{o}}^{2}}.  \label{23}
\end{equation}

\begin{figure}[H]
\begin{minipage}[t]{0.33\textwidth}
        \centering
        \includegraphics[width=\textwidth]{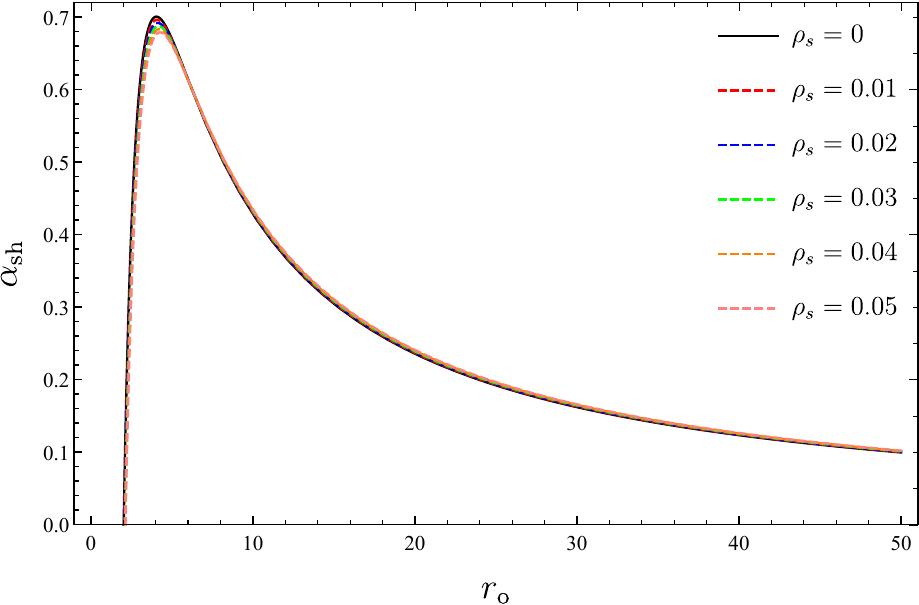}
       \subcaption{ $\omega_{q}=-0.4$.}\label{fig:al1}
   \end{minipage}%
\begin{minipage}[t]{0.33\textwidth}
       \centering
        \includegraphics[width=\textwidth]{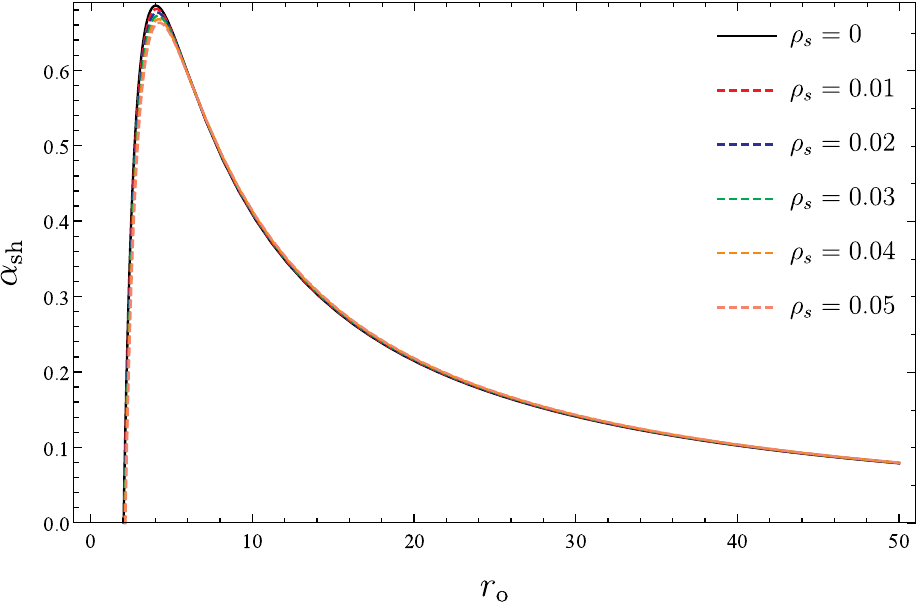}
       \subcaption{ $\omega_{q}=-0.6$}\label{fig:al2}
   \end{minipage}
\begin{minipage}[t]{0.33\textwidth}
       \centering
        \includegraphics[width=\textwidth]{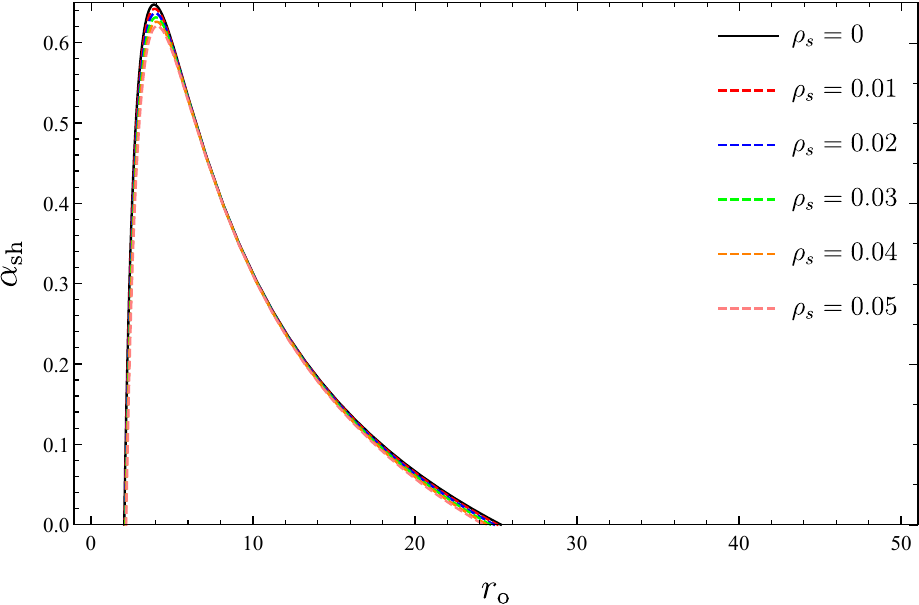}
       \subcaption{ $\omega_{q}=-0.8$}\label{fig:al3}
   \end{minipage}   
\caption{The  angular radius of the shadow with $\alpha_{\mathrm{sh}}$ vs. $r_{\mathrm{o}}$ for different values $\rho_{s}$}
\label{figalpha}
\end{figure}

\begin{figure}[H]
\begin{minipage}[t]{0.32\textwidth}
        \centering
        \includegraphics[width=\textwidth]{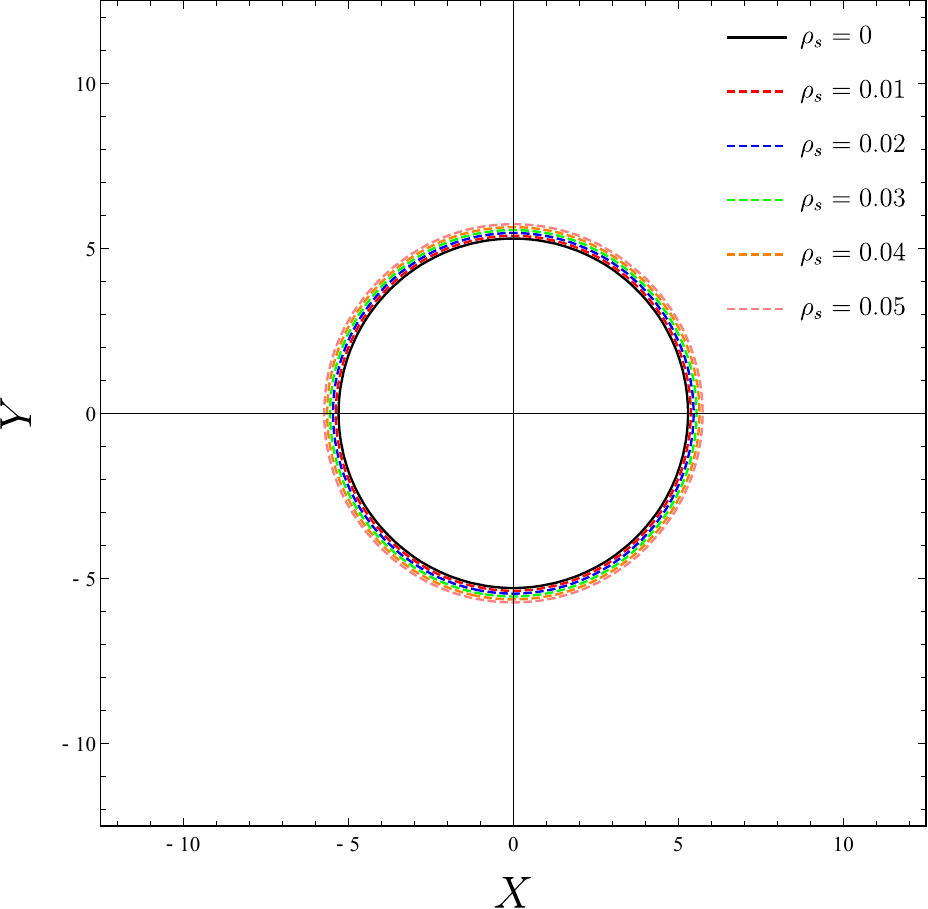}
       \subcaption{ $ \omega_{q}=-0.4$.}\label{fig:xy1}
   \end{minipage}%
\begin{minipage}[t]{0.32\textwidth}
       \centering
        \includegraphics[width=\textwidth]{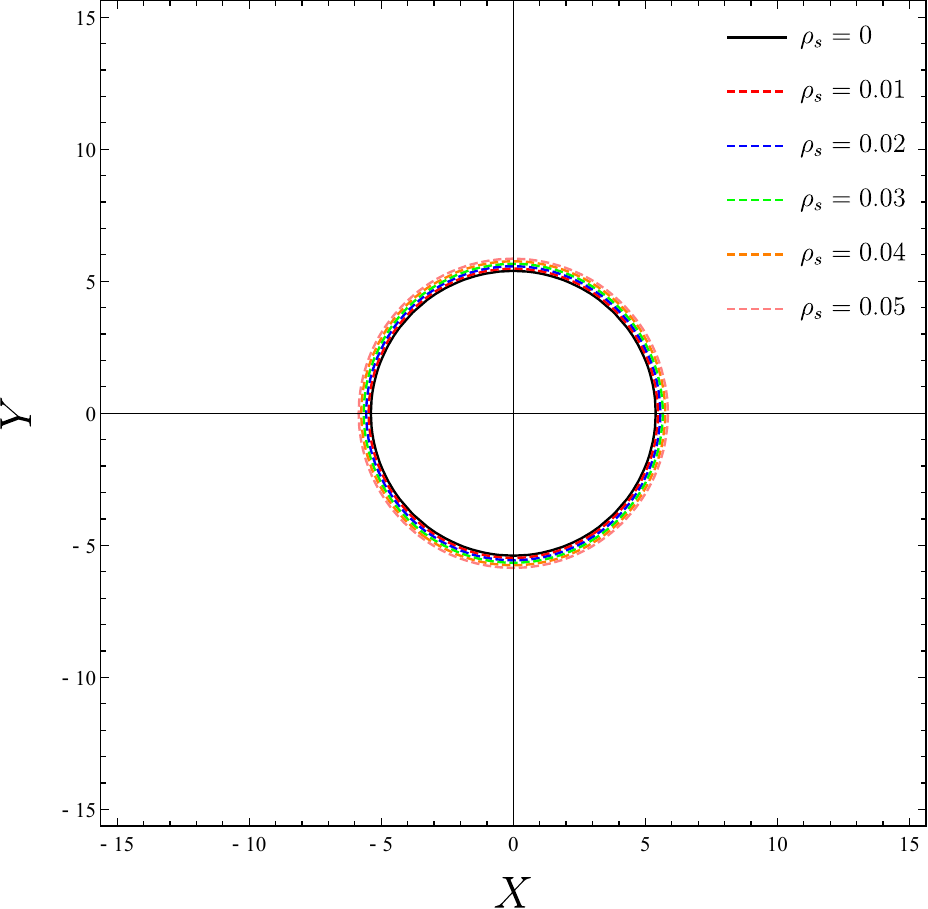}
       \subcaption{ $ \omega_{q}=-0.6$}\label{fig:xy2}
   \end{minipage}
\begin{minipage}[t]{0.32\textwidth}
        \centering
        \includegraphics[width=\textwidth]{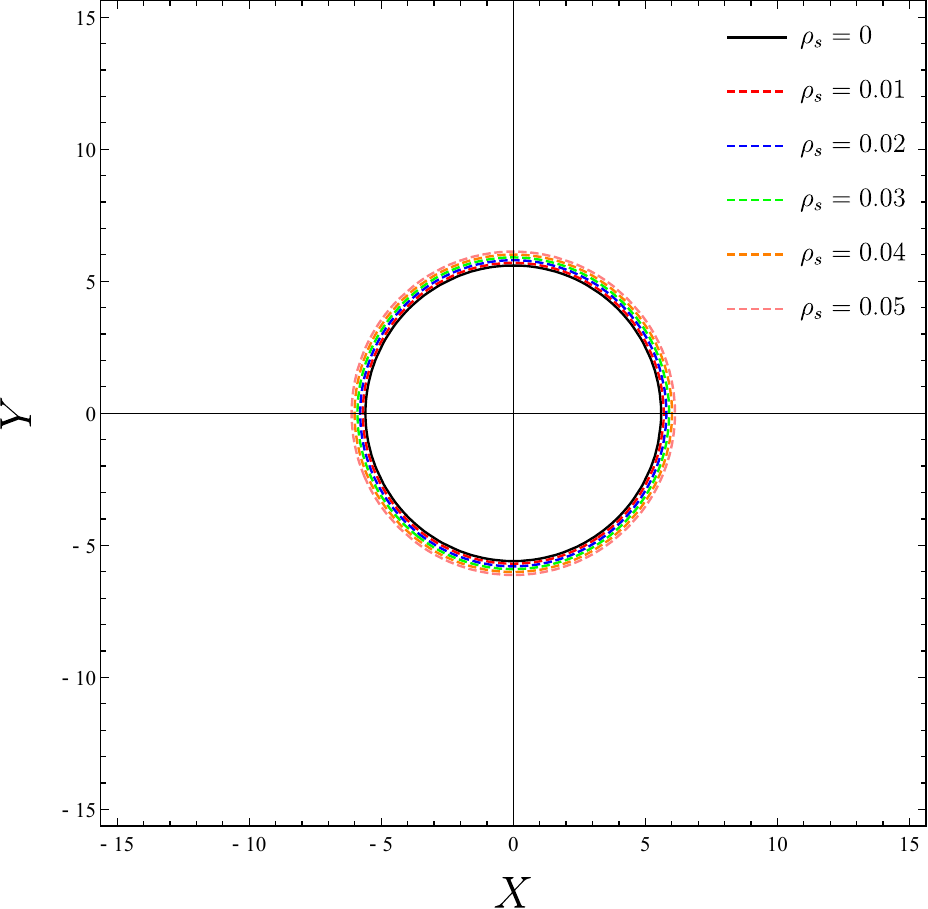}
         \subcaption{ $ \omega_{q}=-0.8$.}\label{fig:xy3}
   \end{minipage}
\caption{The impact of the DM on the shadows of the Schwarzschild BH surrounded by the Dehnen-type DM halo for various values of the DM halo parameters $\rho_{s}$ with $M=1$, $c=0.01$, $r_{s}=0.1$}
\label{figshadow1}
\end{figure}

\begin{figure}[H]
\begin{minipage}[t]{0.32\textwidth}
        \centering
        \includegraphics[width=\textwidth]{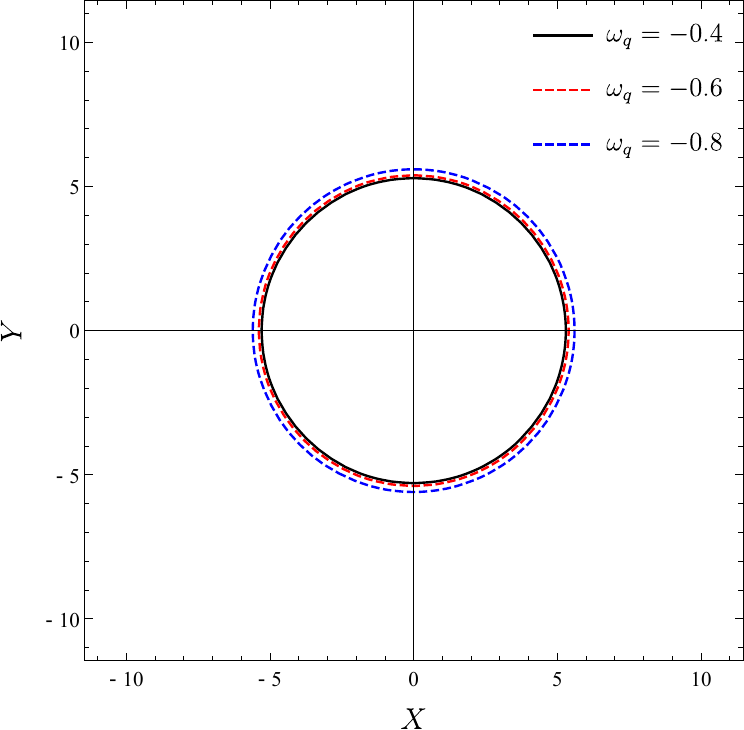}
       \subcaption{ $ \rho_{q}=0.01$.}\label{fig:xy11}
   \end{minipage}%
\begin{minipage}[t]{0.32\textwidth}
       \centering
        \includegraphics[width=\textwidth]{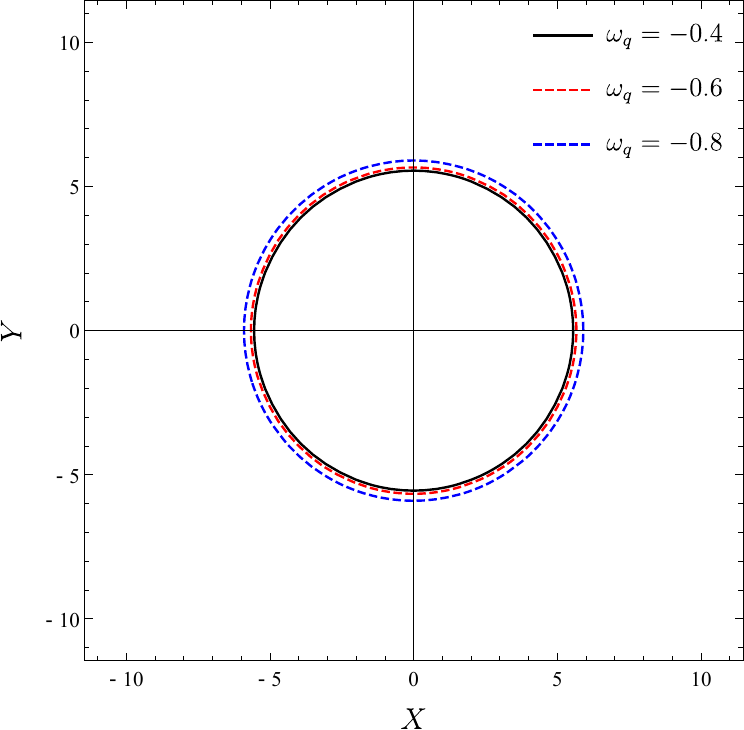}
       \subcaption{ $ \rho_{s}=0.03$}\label{fig:xy22}
   \end{minipage}
\begin{minipage}[t]{0.32\textwidth}
        \centering
        \includegraphics[width=\textwidth]{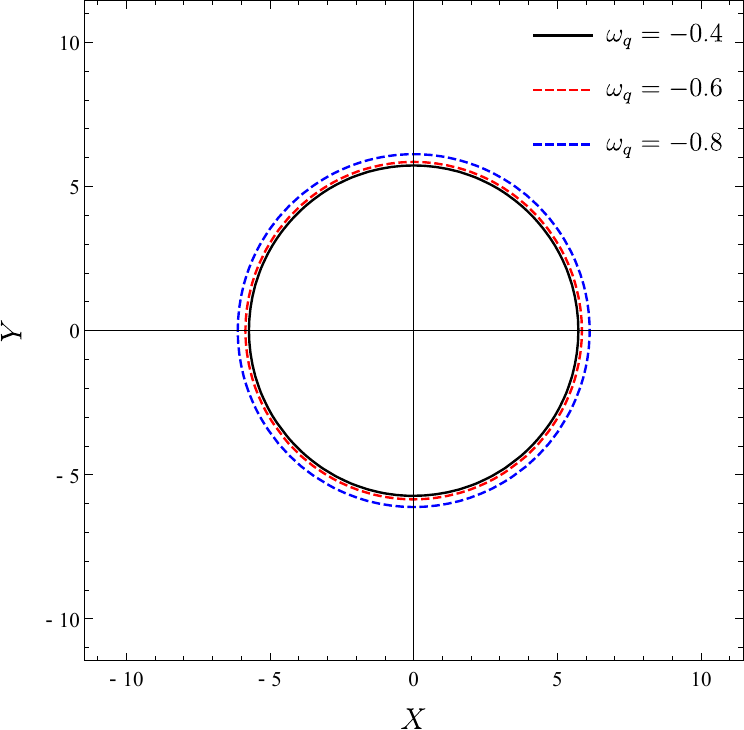}
         \subcaption{ $ \rho_{s}=0.05$.}\label{fig:xy33}
   \end{minipage}
\caption{The impact of the quintessence field on the shadows of the Schwarzschild BH surrounded by the Dehnen-type DM halo for various values of the quintessence field parameters $\omega_{q}$ with $M=1$, $c=0.01$, $r_{s}=0.1$}
\label{figshadow2}
\end{figure}

When the observer is at the photon sphere, where $r_{\mathrm{o}}=$ $r_{\mathrm{c}}$, the angular size of the shadow becomess $\alpha _{\mathrm{sh}}=\pi /2$, meaning the shadow covers exactly half of the observer's sky. Figure \ref{figalpha} illustrates how the angular radius of the shadow, $\alpha _{\mathrm{sh}}$ vs. $r_{\mathrm{o}}$. The graph shows that the angular radius increases with $r_{\mathrm{o}}$, reaches a maximum, and then decreases as $r_{\mathrm{o}}$ continues to grow. This behavior suggests that the shadow's size peaks at a particular radial distance. The presence of DM, $\rho _{s}$, influences the position and magnitude of this maximum, highlighting the impact of DM on the observed shadow, while the presence of the quintessence field affects the maximum angular size of the
shadow.
Now we introduce celestial coordinates, $X$ and $Y$,  to represent the
actual shadow of the BH as observed from the observer's frame%
\begin{equation}
X=\lim_{r_{\mathrm{o}}\rightarrow \infty }\left( -r_{\mathrm{o}}^{2}\sin
\theta _{\mathrm{o}}\frac{d\varphi }{dr}\right) ,
\end{equation}%
\begin{equation}
Y=\lim_{r_{\mathrm{o}}\rightarrow \infty }\left( r_{\mathrm{o}}^{2}\frac{%
d\theta }{dr}\right) ,
\end{equation}

for a static observer at large distance, i.e. at  $r_{\mathrm{o}}\rightarrow
\infty $ in the equatorial plane $\theta _{\mathrm{o}}=\pi /2$, the celestial coordinates simplify to 
\begin{equation}
X^{2}+Y^{2}=b_{\mathrm{c}}^{2}=R_{\mathrm{sh}}^{2}.
\end{equation}%
where $R_{\mathrm{sh}}$  is the radius of the shadow. Now, we examine the influence of the Dehnen-type DM halo and quintessence matter field on the BH shadows. In Figure \ref{figshadow1}, we show the effects of the Dehnen-type DM halo alongside three distinct quintessence matter fields. It is observed that for smaller DM density profile parameter, $\rho_{s}$, the shadows have small radii. However, these graphs do not allow us to clearly assess the impact of the quintessence matter field for a constant DM density profile parameter, $\rho_{s}$. To address this, we present Figure \ref{figshadow2}. We observe that an increase in the quintessence parameter leads to a decrease in the shadow radius.
\section{Weak deflection angle} \label{sec:shadow}
In this section, we examine the behaviors of the deflection angle of
Schwarzschild BHs embedded in a Dehnen-type DM halo with quintessence field by analysing the effect of the central halo density radius $\rho _{s}$ and $r_{s}$. To analyze this behavior, we employ the Gauss-Bonnet theorem. Notably, the deflection angle can be determined using the following relation%

\begin{equation}
\Theta =-\int_{0}^{\pi }\int_{\frac{b}{\sin \varphi }}^{\infty }KdS,
\label{teta}
\end{equation}

where $b$ is the impact parameter, $K$ represents the Gaussian curvature, and $dS$ corresponds to the surface area element of the associated optical metric,

\begin{equation}
d\sigma ^{2}=\frac{dr^{2}}{f\left( r\right) ^{2}}+\frac{r^{2}}{f\left(
r\right) }\,d\varphi ^{2}.
\end{equation}%
with
\begin{equation}
f\left( r\right) =1-\frac{2\,M}{r}-32\pi \rho _{s}r_{s}^{3}\sqrt{\frac{%
r+r_{s}}{r_{s}^{2}\,r}}-c,
\end{equation}%
and $\omega _{q}=-1/3$, which refers to the frustrated network of cosmic strings. Here, the surface area element $dS$ of the associated optical metric is expressed

\begin{equation}
dS\simeq r\left( 1+\frac{3\,M}{r}+\frac{3c}{2}+48\pi \rho
_{s}r_{s}^{2}+24\pi \rho _{s}\frac{r_{s}^{3}}{r}\right) drd\varphi .
\label{dssur}
\end{equation}

The non-zero Christoffel symbols of the above optical metric are calculated as

\begin{equation*}
\Gamma _{rr}^{r}=-\frac{1}{f\left( r\right) }\frac{\partial f\left( r\right) 
}{\partial r},
\end{equation*}%
\begin{equation*}
\Gamma _{\varphi \varphi }^{r}=-\frac{r}{2}\left( r\frac{\partial f\left(
r\right) }{\partial r}-2f\left( r\right) \right) ,
\end{equation*}%
\begin{equation*}
\Gamma _{r\varphi }^{\varphi }=\Gamma _{\varphi r}^{\varphi }=\frac{1}{r}-%
\frac{1}{2f\left( r\right) }\frac{\partial f\left( r\right)}{\partial r},
\end{equation*}
and the Ricci scalar of the optical metric computes as:%
\begin{equation}
\mathbf{R}=f\left( r\right) \frac{\partial ^{2}f\left( r\right) }{\partial r^{2}}-\frac{1}{2}\left( \frac{\partial f\left( r\right) }{\partial r}\right)^{2}.
\end{equation}

The Gaussian curvature of the optical metric can be defined as%
\begin{equation}
K=\frac{\mathbf{R}}{2}.
\end{equation}%
For the sake of simplicity, we consider the Gaussian optical curvature up to the leading order terms $\mathcal{O}\left( M^{2},c^{2},\rho_{s}^{2},r_{s}^{4}\right) $, expressed as%

\begin{equation}
K=-\frac{2M}{r^{3}}-\frac{16\pi \rho _{s}r_{s}^{3}}{r^{3}}+\frac{56\pi \rho_{s}r_{s}^{2}M}{r^{3}}+\frac{48\pi \rho _{s}r_{s}^{3}M}{r^{4}}+\frac{2cM}{r^{3}}+\frac{16\pi \rho _{s}r_{s}^{3}c}{r^{3}}.  \label{curvature}
\end{equation}

By substituting the surface area element from Eq. (\ref{dssur}) and the leading-order terms of the Gaussian curvature from Eq. (\ref{curvature}) into Eq. (\ref{teta}), the deflection angle is computed as follows:
\begin{equation}
\Theta =\frac{4M}{b}+\frac{32\pi \rho _{s}r_{s}^{3}}{b}+\frac{2cM}{b}+\frac{80\pi \rho _{s}r_{s}^{2}M\left( 1-\frac{9}{2}\right) }{b}+\frac{6\pi^{2}\rho _{s}r_{s}^{3}M\left( 1-6c\right) }{b^{2}}+\frac{16\pi \rho_{s}r_{s}^{3}c}{b}.
\end{equation}

\begin{figure}[H]
\begin{minipage}[t]{0.5\textwidth}
        \centering
        \includegraphics[width=\textwidth]{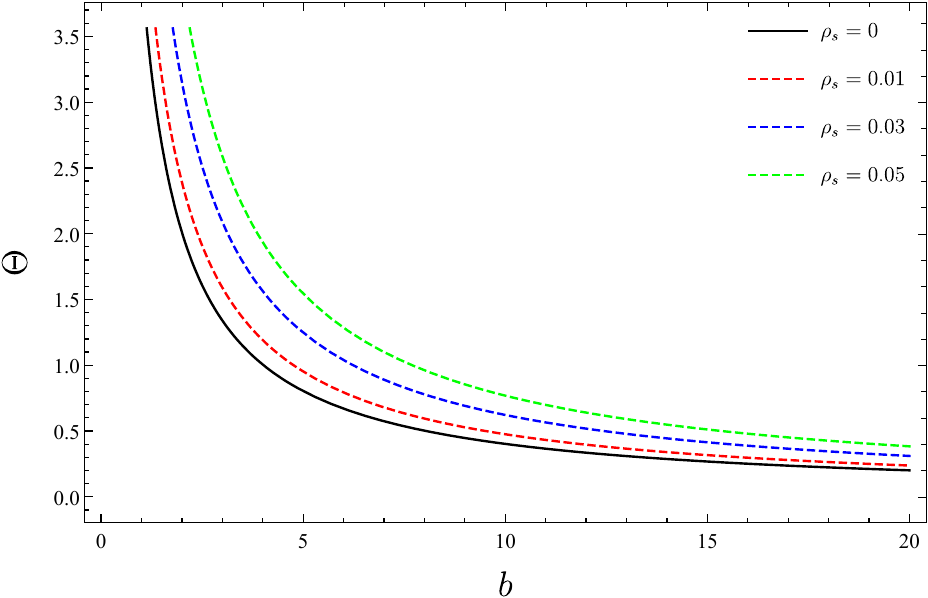}
       \subcaption{ $r_{q}=0.5$.}\label{fig:def1}
   \end{minipage}%
\begin{minipage}[t]{0.5\textwidth}
       \centering
        \includegraphics[width=\textwidth]{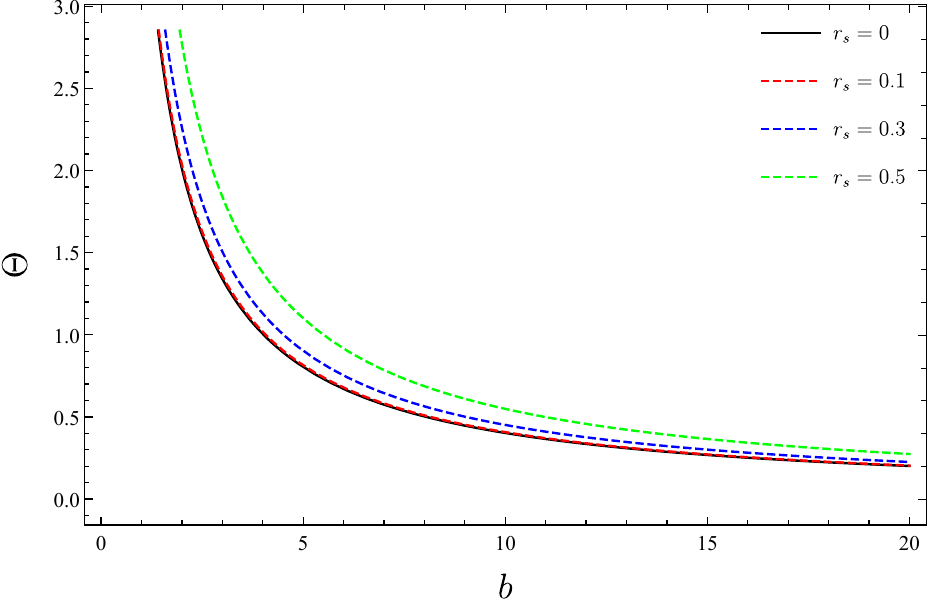}
       \subcaption{ $\rho_{s}=0.02$}\label{fig:adef2}
   \end{minipage}
\caption{Deflection angle vs impact parameter $b$ for $M=1$ and $c=0.01$}
\label{figdeflection1}
\end{figure}

Having computed the deflection angle, we proceed to examine and explore the related behavior. Specifically, we investigate the influence of the impact parameter $b$, the core density $\rho _{s}$, the radius $r_{s}$, and the quintessence parameter $c$.
\begin{itemize}
\item  Fig.\ref{figdeflection1} shows the behavior of $\Theta $ vs.the impact parameter $b$
\end{itemize}

\begin{enumerate}
\item In figure \ref{fig:def1}, the plot shows the behavior of $\Theta $  with respect to $b$ for fixed $r_{s},c$ and $M$ while varying $\rho _{s}$. We observed that the deflection angle increases for smaller $b$ and higher $\rho_{s}$. The figure illustrates that denser halos produce stronger deflection. 
\item In figure \ref{fig:adef2}, the plot shows deflection angle versus impact parameter $b$ for fixed $\rho _{s},c$ and $M$ for different core radii $r_{s}$. The deflection is weakest for $r_{s}=0$,  and becomes stronger as $r_{s}$ increases, demonstrating how the core size influences gravitational lensing. The curves converge at large $b$ where the influence of the halo diminishes, and diverge at small $b$.
\end{enumerate}

\begin{figure}[H]
\begin{minipage}[t]{0.5\textwidth}
        \centering
        \includegraphics[width=\textwidth]{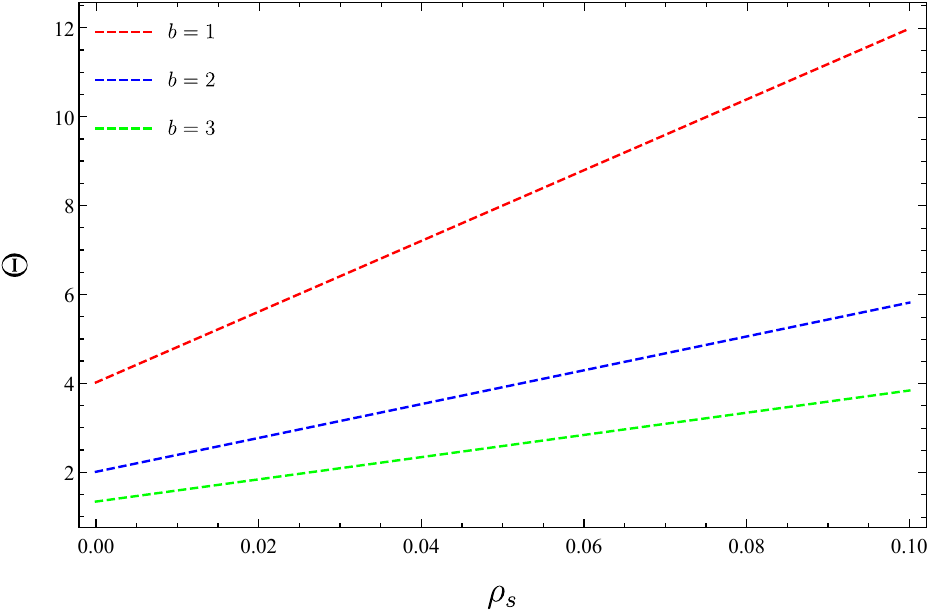}
       \subcaption{ $r_{q}=0.5$.}\label{fig:def3}
   \end{minipage}%
\begin{minipage}[t]{0.5\textwidth}
       \centering
        \includegraphics[width=\textwidth]{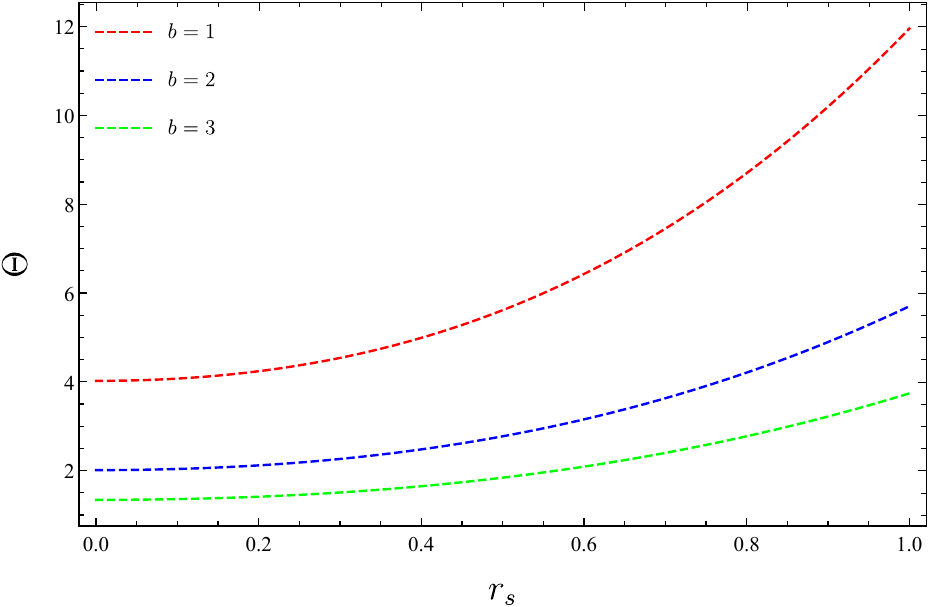}
       \subcaption{ $\rho_{s}=0.02$}\label{fig:adef4}
   \end{minipage}
\caption{Deflection angle for varying values of $\rho_{s}$ by setting $r_{q}=0.5$ (left) and for varying $r_{s}$ by setting $\rho_{s}=0.02$ with $M=1$ and $c=0.01$}
\label{figdeflection2}
\end{figure}

\begin{itemize}
\item  Fig.\ref{figdeflection2} shows the behavior of $\Theta $ as a function of the core density $\rho_{s}$ (left) and radius $r_{s}$ (right)$r_{s}$ for multiple values of the impact parameter $b$
\end{itemize}
\begin{enumerate}
\item The fig. \ref{fig:def3} shows how the deflection angle varies with core density $\rho_{s}$ for different impact parameters.  As $\rho_{s}$ increases, the deflection angle rises, indicating that denser cores cause stronger deflection. Smaller impact parameters result in larger deflection angles. 
\item The fig. \ref{fig:adef4} illustrates the relationship between the deflection angle and the radius $r_{s}$  for impact parameters $b=1,2,3$. As $r_{s}$ increases, the deflection angle rises, peaking at higher radii.
\end{enumerate}
\begin{itemize}
\item  Fig.\ref{figdeflection2} shows the behavior of $\Theta $ as a function of the quintessence parameter $c$.
\end{itemize}
\begin{enumerate}
\item The Fig. \ref{fig:def3} shows that the deflection angle decreases as the quintessence parameter increases, indicating an inverse relationship. Higher core densities result in larger deflection angles. This suggests that both quintessence and core density significantly influence gravitational lensing.
\item The Fig.\ref{fig:adef4} plots the deflection angle against the quintessence parameter for different radii $r_{s}$. The deflection angle decreases as quintessence increases, showing an inverse relationship. Larger radii result in smaller deflection angles, suggesting that greater values of $r_{s}$ produce weaker lensing effects.
\end{enumerate}
\begin{figure}[H]
\begin{minipage}[t]{0.5\textwidth}
        \centering
        \includegraphics[width=\textwidth]{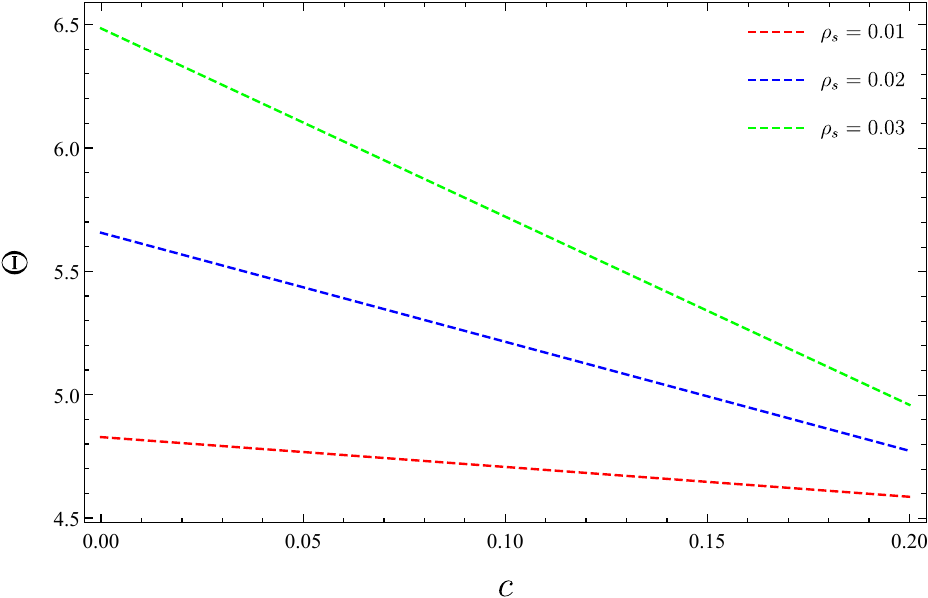}
       \subcaption{ $r_{q}=0.5$.}\label{fig:def5}
   \end{minipage}%
\begin{minipage}[t]{0.5\textwidth}
       \centering
        \includegraphics[width=\textwidth]{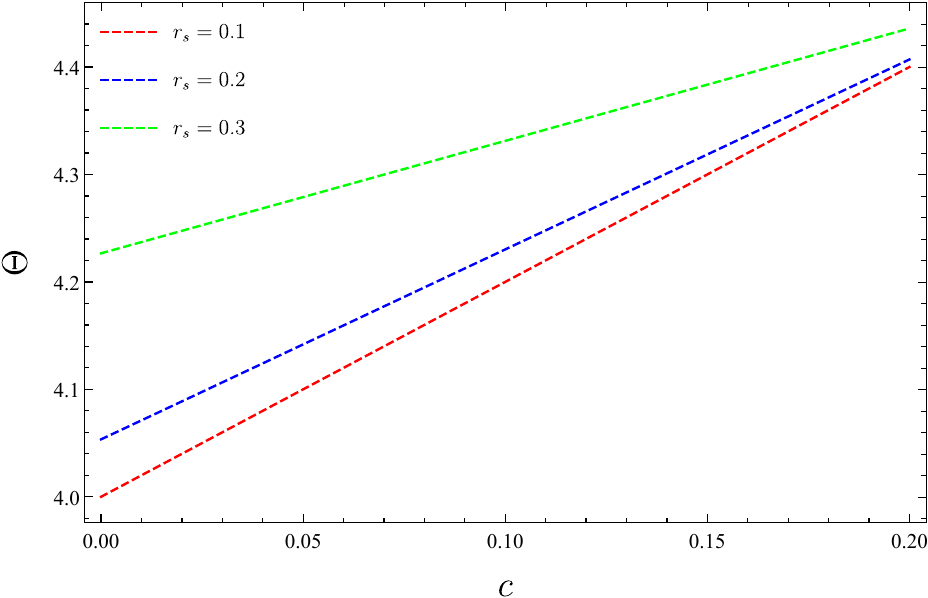}
       \subcaption{ $\rho_{s}=0.02$}\label{fig:adef6}
   \end{minipage}
\caption{Deflection angle vs quintessence parameter $b$ for $M=1$ and $c=0.01$ }
\label{figdeflection3}
\end{figure}
\section{Scalar Perturbations} \label{sec:qnm}

In this part, we look at the dynamics of a massless scalar field in the context of a specific Schwarzschild-like BH solution with a Dehnen-type DM halo. We begin by explicitly constructing the massless Klein-Gordon equation, which determines the evolution of the scalar field in the given space-time geometry.

The massless scalar field wave equation is described by the Klein-Gordon equation as follows:
\begin{equation}
\frac{1}{\sqrt{-g}}\,\partial_{\mu}\left[\left(\sqrt{-g}\,g^{\mu\nu}\,\partial_{\nu}\right)\,\Psi\right]=0,\label{ff1}    
\end{equation}
where $\Psi$ is the wave function of the scalar field, $g_{\mu\nu}$ is the covariant metric tensor, $g=\det(g_{\mu\nu})$ is the determinant of the metric tensor, $g^{\mu\nu}$ is the contrvariant form of the metric tensor, and $\partial_{\mu}$ is the partial derivative with respect to the coordinate systems.

Using the following coordinate transformation i.e. tortoise coordinate 
\begin{eqnarray}
    dr_*=\frac{dr}{\mathcal{F}(r)}\label{ff2}
\end{eqnarray}
into the line-element Eq. (\ref{bb1}) results
\begin{equation}
    ds^2=\mathcal{F}(r_*)\,\left(-dt^2+dr^2_{*}\right)+\mathcal{H}^2(r_*)\,\left(d\theta^2+\sin^2 \theta\,d\phi^2\right),\label{ff3}
\end{equation}
where $\mathcal{F}(r_*)$ and $\mathcal{H}(r_*)$ are the new metric functions of $r_*$. 

Let us consider the following scalar field wave function ansatz form
\begin{equation}
    \Psi(t, r_{*},\theta, \phi)=\exp(i\,\omega\,t)\,Y^{m}_{\ell} (\theta,\phi)\,\frac{\psi(r_*)}{r_{*}},\label{ff4}
\end{equation}
where $\omega$ is (possibly complex) the temporal frequency, $\psi (r)$ is a propagating scalar field in the candidate space-time, and $Y^{m}_{\ell} (\theta,\phi)$ is the spherical harmonics. 

We can write the wave equation (\ref{ff1}) as follows:
\begin{equation}
    \frac{\partial^2 \psi(r_*)}{\partial r^2_{*}}+\left(\omega^2-V_s\right)\,\psi(r_*)=0,\label{ff5}
\end{equation}
where the perturbative potential $V_s$ is given by
\begin{eqnarray}
V_s=\left(\frac{\ell\,(\ell+1)}{r^2}+\frac{\mathcal{F}'(r)}{r}\right)\,\mathcal{F}(r).\label{ff6}
\end{eqnarray}

\begin{figure}[ht!]
    \centering
    \includegraphics[width=0.45\linewidth]{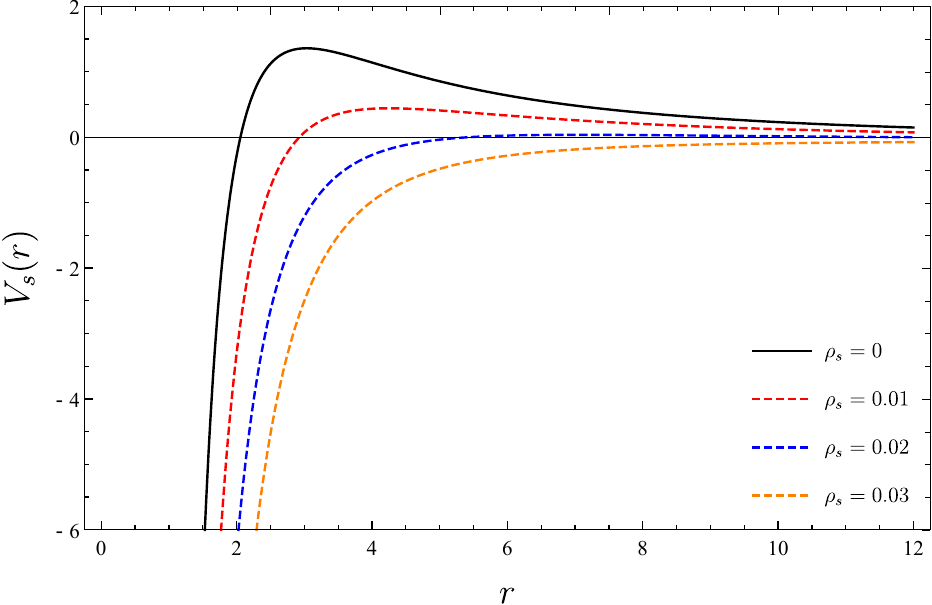}\quad\quad
    \includegraphics[width=0.45\linewidth]{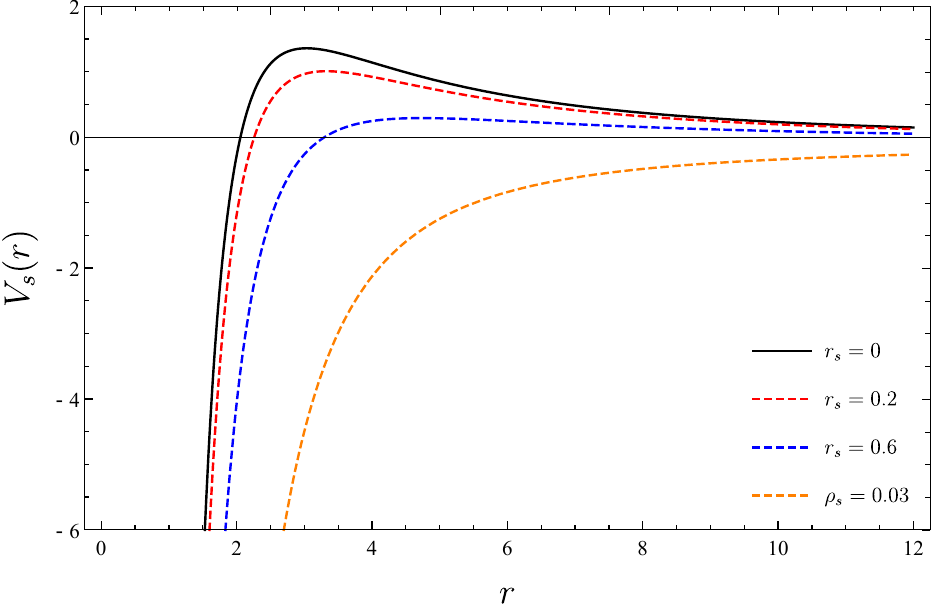}
    \caption{Scalar perturbative potential $V_s$ as a function of $r$. Left panel: for varying values of $\rho_s$ by setting $r_s=0.5$, right panel: for varying values of $r_s$ by setting $\rho_s=0.02$. Here, $M=1$ and $c=0.01$.}
    \label{potScalar}
\end{figure}

Inserting the metric function (\ref{bb2}), Eq. (\ref{ff6}) becomes 
\begin{eqnarray}
V_s=\left[\frac{\ell\,(\ell+1)}{r^2}+\frac{2\,M}{r^3}-\frac{16\pi r_s^3 \rho_s}{r^3 \sqrt{\frac{r+r_s}{r_s^2\,r}}}-\frac{c(1+3w)}{r^{-(2+3w)}}\right]\left(1-\frac{2\,M}{r}-32\pi \rho _{s}r_{s}^{3}\sqrt{\frac{r+r_s}{r_s^2\,r}}-\frac{c}{r^{3\,w+1}}\right).\label{ff7}
\end{eqnarray}
In Figure \ref{potScalar}, we display the scalar perturbative potential $V_s$ given in Eq. (\ref{ff7}) as a function of $r$ for varied values of the core density and radius parameters, while maintaining the state parameter $w= -0.8$. We notice that as both $\rho_s$ and $r_s$ grow, the perturbative potential diminishes and the peak shifts right.

\subsection{Quasinormal modes}

Quasinormal modes govern the late-time decay of signal in the vicinity of a BH \cite{Konoplya:2011qq, Kokkotas:1999bd, Bolokhov:2025uxz}. They are the characteristic oscillation frequencies corresponding to purely ingoing boundary conditions at the event horizon and purely outgoing conditions at infinity or at the cosmological horizon in de Sitter spacetime:
\[
\Psi(r) \sim 
\begin{cases}
e^{-i \omega r_*}, & r \to r_h \quad (\text{event horizon}), \\
e^{+i \omega r_*}, & r \to \infty \quad (\text{asymptotically flat}), \\
e^{+i \omega r_*}, & r \to r_c \quad (\text{de Sitter horizon}).
\end{cases}
\]
Here, \( r_* \) is the tortoise coordinate, \( \omega \) is the complex quasinormal frequency, \( r_h \) denotes the event horizon radius, and \( r_c \) is the radius of the cosmological horizon (when present). When $3 \omega_q +1 <0$ the boundary conditions are the same as for asymptotically de Sitter ones.

In this work, we employ two complementary approaches: the higher-order WKB method with Padé approximants \cite{Iyer:1986np, Konoplya:2003ii, Matyjasek:2017psv, Konoplya:2019hlu} and time-domain integration \cite{Gundlach:1993tp}. Both methods are typically sufficiently accurate for computing the low-lying quasinormal modes but lose precision when applied to higher overtones as was demonstrated in numerous works \cite{Konoplya:2005sy, Konoplya:2001ji, Gong:2023ghh, Bolokhov:2024ixe, Bolokhov:2023dxq, Bolokhov:2023ruj, Skvortsova:2024wly, Skvortsova:2024eqi, Skvortsova:2023zmj, Churilova:2021tgn, Churilova:2019qph, Dubinsky:2024aeu, Malik:2024elk}.  Since our focus is on the modes that dominate the ringdown phase—those with potential observational relevance—we restrict our analysis to the least damped fundamental mode. Nevertheless, overtones can carry valuable information about the near-horizon geometry \cite{Konoplya:2022pbc} and warrant a dedicated investigation using more accurate and sophisticated techniques, such as the Leaver method \cite{Leaver:1985ax}.

The Wentzel–Kramers–Brillouin (WKB) method is a semi-analytic technique widely used to compute quasinormal modes of BHs, especially in the regime of high multipole numbers $\ell \geq n$. It is based on matching WKB expansions across the turning points of an effective potential barrier derived from the wave equation. The quasinormal frequencies are found from the condition
\begin{equation}
\frac{i Q_0}{\sqrt{2 Q_0''}} - \sum_{j=2}^{N} \Lambda_j = n + \frac{1}{2},
\end{equation}
where $Q_0 =V_{0} -\omega^2$ and $V_{0}$ is the potential at its maximum, $Q_0''$ is the second derivative at the maximum, $\Lambda_j$ are higher-order correction terms, and $n$ is the overtone number. The use of Padé approximants to resum the WKB series improves its accuracy, especially for low $\ell$ and $n$~\cite{Konoplya:2019hlu}.

The time-domain integration method involves solving the perturbation equation numerically in the time domain, often using light-cone coordinates. A common approach is to rewrite the wave equation
\begin{equation}
\left( \frac{\partial^2}{\partial t^2} - \frac{\partial^2}{\partial r_*^2} + V(r) \right) \Psi(t, r) = 0
\end{equation}
in terms of the null coordinates $u = t - r_*$ and $v = t + r_*$, and evolve the field $\Psi(u,v)$ using finite difference schemes suggested in \cite{Gundlach:1993tp}. The resulting time-domain profile reveals damped oscillations, from which the dominant quasinormal modes can be extracted via Prony method. This technique is particularly useful for capturing the full dynamical behavior of perturbations, including the late-time tails and nonlinear effects.

\begin{table*}
\centering
\begin{tabular}{c c c c c}
\hline
\hline
$\omega_q$ & $\rho_s$ & WKB6 $m=3$ & WKB7 $m=3$ & difference ($\%$)  \\
\hline
$-0.35$ & $0.05$ & $0.109396-0.102286 i$ & $0.109382-0.102133 i$ & $0.103\%$\\
$-0.35$ & $0.1$ & $0.109242-0.102190 i$ & $0.109228-0.102037 i$ & $0.103\%$\\
$-0.35$ & $0.2$ & $0.108935-0.101998 i$ & $0.108921-0.101844 i$ & $0.104\%$\\
$-0.35$ & $0.3$ & $0.108632-0.101808 i$ & $0.108618-0.101653 i$ & $0.104\%$\\
$-0.35$ & $0.4$ & $0.108331-0.101618 i$ & $0.108316-0.101463 i$ & $0.105\%$\\
$-0.35$ & $0.5$ & $0.108032-0.101429 i$ & $0.108018-0.101274 i$ & $0.105\%$\\
$-0.5$ & $0.05$ & $0.107185-0.101176 i$ & $0.107153-0.100993 i$ & $0.126\%$\\
$-0.5$ & $0.1$ & $0.107031-0.101079 i$ & $0.106999-0.100896 i$ & $0.126\%$\\
$-0.5$ & $0.2$ & $0.106725-0.100885 i$ & $0.106692-0.100702 i$ & $0.126\%$\\
$-0.5$ & $0.3$ & $0.106421-0.100693 i$ & $0.106389-0.100509 i$ & $0.127\%$\\
$-0.5$ & $0.4$ & $0.106121-0.100501 i$ & $0.106088-0.100317 i$ & $0.128\%$\\
$-0.5$ & $0.5$ & $0.105823-0.100310 i$ & $0.105790-0.100126 i$ & $0.128\%$\\
$-0.75$ & $0.05$ & $0.100198-0.100251 i$ & $0.100180-0.100293 i$ & $0.0321\%$\\
$-0.75$ & $0.1$ & $0.100036-0.100149 i$ & $0.100019-0.100191 i$ & $0.0323\%$\\
$-0.75$ & $0.2$ & $0.099716-0.099944 i$ & $0.099699-0.099987 i$ & $0.0328\%$\\
$-0.75$ & $0.3$ & $0.099399-0.099741 i$ & $0.099381-0.099784 i$ & $0.0333\%$\\
$-0.75$ & $0.4$ & $0.099084-0.099538 i$ & $0.099066-0.099582 i$ & $0.0338\%$\\
$-0.75$ & $0.5$ & $0.098772-0.099336 i$ & $0.098754-0.099381 i$ & $0.0342\%$\\
\hline
\hline
\end{tabular}
\caption{Quasinormal modes of the $\ell = 0$ scalar field for $r_s = 0.1$, $c = 0.01$, and $M = 1$, calculated using the WKB method at different orders and with Padé approximants. 
}\label{QNMtable1}
\end{table*}

\begin{table*}
\centering
\begin{tabular}{c c c c c}
\hline
\hline
$\omega_q$ & $\rho_s$ & WKB6 $m=3$ & WKB7 $m=3$ & difference ($\%$)  \\
\hline
$-0.35$ & $0.05$ & $0.287650-0.095488 i$ & $0.287653-0.095487 i$ & $0.0009\%$\\
$-0.35$ & $0.1$ & $0.287230-0.095392 i$ & $0.287233-0.095391 i$ & $0.001\%$\\
$-0.35$ & $0.2$ & $0.286395-0.095201 i$ & $0.286398-0.095200 i$ & $0.0010\%$\\
$-0.35$ & $0.3$ & $0.285568-0.095011 i$ & $0.285571-0.095010 i$ & $0.0011\%$\\
$-0.35$ & $0.4$ & $0.284749-0.094822 i$ & $0.284752-0.094821 i$ & $0.0011\%$\\
$-0.35$ & $0.5$ & $0.283936-0.094634 i$ & $0.283940-0.094633 i$ & $0.0012\%$\\
$-0.5$ & $0.05$ & $0.284213-0.094123 i$ & $0.284216-0.094122 i$ & $0.0009\%$\\
$-0.5$ & $0.1$ & $0.283793-0.094027 i$ & $0.283795-0.094026 i$ & $0.001\%$\\
$-0.5$ & $0.2$ & $0.282958-0.093834 i$ & $0.282961-0.093834 i$ & $0.0010\%$\\
$-0.5$ & $0.3$ & $0.282131-0.093643 i$ & $0.282134-0.093642 i$ & $0.0011\%$\\
$-0.5$ & $0.4$ & $0.281311-0.093453 i$ & $0.281314-0.093452 i$ & $0.0011\%$\\
$-0.5$ & $0.5$ & $0.280498-0.093264 i$ & $0.280501-0.093263 i$ & $0.0012\%$\\
$-0.75$ & $0.05$ & $0.272071-0.090755 i$ & $0.272073-0.090755 i$ & $0.0010\%$\\
$-0.75$ & $0.1$ & $0.271636-0.090652 i$ & $0.271639-0.090652 i$ & $0.0010\%$\\
$-0.75$ & $0.2$ & $0.270773-0.090447 i$ & $0.270776-0.090446 i$ & $0.0011\%$\\
$-0.75$ & $0.3$ & $0.269918-0.090243 i$ & $0.269921-0.090242 i$ & $0.0011\%$\\
$-0.75$ & $0.4$ & $0.269070-0.090040 i$ & $0.269074-0.090039 i$ & $0.0012\%$\\
$-0.75$ & $0.5$ & $0.268230-0.089838 i$ & $0.268234-0.089837 i$ & $0.0012\%$\\
\hline
\hline
\end{tabular}
\caption{Quasinormal modes of the $\ell = 1$ scalar field for $r_s = 0.1$, $c = 0.01$, and $M = 1$, calculated using the WKB method at different orders and with Padé approximants. 
}\label{QNMtable2}
\end{table*}

\begin{table*}
\centering
\begin{tabular}{c c c c c}
\hline
\hline
$\omega_q$ & $\rho_s$ & WKB6 $m=3$ & WKB7 $m=3$ & difference ($\%$)  \\
\hline
$-0.35$ & $0.05$ & $0.475162-0.094612 i$ & $0.475162-0.094611 i$ & $0.00014\%$\\
$-0.35$ & $0.1$ & $0.474468-0.094516 i$ & $0.474469-0.094515 i$ & $0.00014\%$\\
$-0.35$ & $0.2$ & $0.473090-0.094325 i$ & $0.473090-0.094324 i$ & $0.00014\%$\\
$-0.35$ & $0.3$ & $0.471724-0.094135 i$ & $0.471725-0.094135 i$ & $0.00014\%$\\
$-0.35$ & $0.4$ & $0.470371-0.093946 i$ & $0.470371-0.093946 i$ & $0.00014\%$\\
$-0.35$ & $0.5$ & $0.469029-0.093758 i$ & $0.469030-0.093758 i$ & $0.00015\%$\\
$-0.5$ & $0.05$ & $0.469938-0.093226 i$ & $0.469938-0.093226 i$ & $0.00014\%$\\
$-0.5$ & $0.1$ & $0.469244-0.093129 i$ & $0.469244-0.093129 i$ & $0.00014\%$\\
$-0.5$ & $0.2$ & $0.467865-0.092937 i$ & $0.467866-0.092937 i$ & $0.00014\%$\\
$-0.5$ & $0.3$ & $0.466499-0.092746 i$ & $0.466499-0.092746 i$ & $0.00014\%$\\
$-0.5$ & $0.4$ & $0.465145-0.092556 i$ & $0.465145-0.092555 i$ & $0.00014\%$\\
$-0.5$ & $0.5$ & $0.463802-0.092367 i$ & $0.463803-0.092366 i$ & $0.00014\%$\\
$-0.75$ & $0.05$ & $0.451895-0.089483 i$ & $0.451896-0.089483 i$ & $0.00013\%$\\
$-0.75$ & $0.1$ & $0.451180-0.089380 i$ & $0.451180-0.089380 i$ & $0.00013\%$\\
$-0.75$ & $0.2$ & $0.449758-0.089174 i$ & $0.449758-0.089174 i$ & $0.00013\%$\\
$-0.75$ & $0.3$ & $0.448349-0.088970 i$ & $0.448349-0.088969 i$ & $0.00013\%$\\
$-0.75$ & $0.4$ & $0.446952-0.088766 i$ & $0.446953-0.088766 i$ & $0.00013\%$\\
$-0.75$ & $0.5$ & $0.445568-0.088564 i$ & $0.445568-0.088564 i$ & $0.00013\%$\\
\hline
\hline
\end{tabular}
\caption{Quasinormal modes of the $\ell = 2$ scalar field for $r_s = 0.1$, $c = 0.01$, and $M = 1$, calculated using the WKB method at different orders and with Padé approximants. 
}\label{QNMtable3}
\end{table*}

\begin{table*}
\centering
\begin{tabular}{c c c c c}
\hline
\hline
$r_s$ & $\rho_s$ & WKB6 $m=3$ & WKB7 $m=3$ & difference ($\%$)  \\
\hline
$0.1$ & $0.05$ & $0.287650-0.095488 i$ & $0.287653-0.095487 i$ & $0.0009\%$\\
$0.1$ & $0.1$ & $0.287230-0.095392 i$ & $0.287233-0.095391 i$ & $0.001\%$\\
$0.1$ & $0.2$ & $0.286395-0.095201 i$ & $0.286398-0.095200 i$ & $0.0010\%$\\
$0.1$ & $0.3$ & $0.285568-0.095011 i$ & $0.285571-0.095010 i$ & $0.0011\%$\\
$0.1$ & $0.4$ & $0.284749-0.094822 i$ & $0.284752-0.094821 i$ & $0.0011\%$\\
$0.1$ & $0.5$ & $0.283936-0.094634 i$ & $0.283940-0.094633 i$ & $0.0012\%$\\
$0.3$ & $0.05$ & $0.277037-0.093066 i$ & $0.277042-0.093065 i$ & $0.0018\%$\\
$0.3$ & $0.1$ & $0.267221-0.090722 i$ & $0.267229-0.090721 i$ & $0.0026\%$\\
$0.3$ & $0.2$ & $0.250416-0.086492 i$ & $0.250426-0.086491 i$ & $0.0040\%$\\
$0.3$ & $0.3$ & $0.236445-0.082778 i$ & $0.236458-0.082779 i$ & $0.0051\%$\\
$0.3$ & $0.4$ & $0.224563-0.079488 i$ & $0.224577-0.079489 i$ & $0.0060\%$\\
$0.3$ & $0.5$ & $0.214280-0.076546 i$ & $0.214295-0.076548 i$ & $0.0068\%$\\
$0.5$ & $0.05$ & $0.245117-0.085172 i$ & $0.245131-0.085173 i$ & $0.0054\%$\\
$0.5$ & $0.1$ & $0.217101-0.077472 i$ & $0.217119-0.077474 i$ & $0.0077\%$\\
$0.5$ & $0.2$ & $0.181147-0.066656 i$ & $0.181166-0.066661 i$ & $0.0103\%$\\
$0.5$ & $0.3$ & $0.158182-0.059249 i$ & $0.158201-0.059254 i$ & $0.0115\%$\\
$0.5$ & $0.4$ & $0.141831-0.053757 i$ & $0.141848-0.053763 i$ & $0.0122\%$\\
$0.5$ & $0.5$ & $0.129410-0.049472 i$ & $0.129427-0.049477 i$ & $0.0127\%$\\
\hline
\hline
\end{tabular}
\caption{Quasinormal modes of the $\ell=1$ scalar perturbations, $\omega_q=-0.35$, $c=0.01$, $M=1$  calculated using the WKB method at different orders and with Padé approximants.}\label{QNMtable4}
\end{table*}

\begin{figure}
\resizebox{\linewidth}{!}{\includegraphics{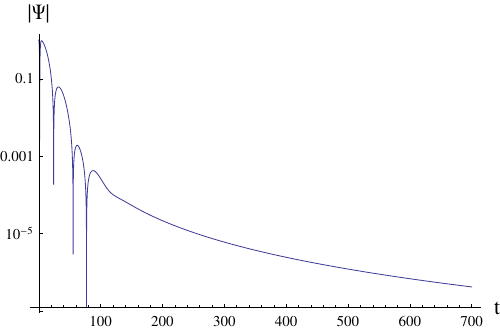}\includegraphics{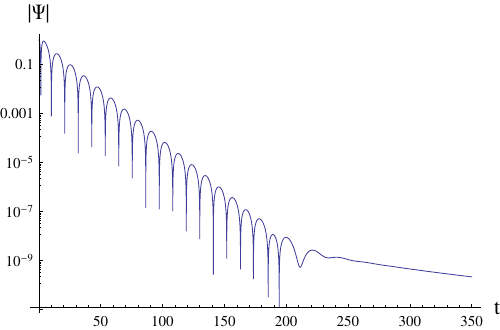}}
\caption{Left panel: Semi-logarithmic time-domain profile for $\ell=0$, $M=1$, $r_s =0.1$, $\rho_s =0.05$, $\omega_q =-0.35$, $c=0.01$. The least damped mode of the Schwarzschild branch is $\omega = - 0.0245 i$, while the least damped mode of the de Sitter-like branch is $\omega = 0.108 - 0.103 i$. Right panel: The same for $\ell=1$: The time domain integration for the Schwarzschild branch gives $\omega = 0.287656 - 0.0954876 i$, which is in perfect agreement with the WKB data $\omega = 0.287650 - 0.095488 i$}\label{fig:TDL0}
\end{figure}

\newpage
When \( 3\omega_q + 1 < 0 \), the spacetime becomes asymptotically de Sitter-like, implying the presence of a cosmological horizon. In particular, when \( \omega_q = -1 \), the exact de Sitter asymptotic behavior is recovered. For asymptotically de Sitter BHs, an additional family of quasinormal modes emerges—those associated with the empty de Sitter spacetime, which become perturbed by the presence of the BH. When the BH radius is much smaller than the de Sitter radius, the quasinormal modes belonging to this de Sitter branch can be approximated by the expression \cite{Konoplya:2022kld}:
\begin{equation}
\omega_n = \omega^{(\text{dS})}_n \left( 1 - \frac{M}{r_c} + \mathcal{O}\left( \left( \frac{M}{r_c} \right)^2 \right) \right),
\end{equation}
where the purely imaginary modes of the empty de Sitter spacetime for massless and massive fields are given by \cite{Lopez-Ortega:2006aal, Lopez-Ortega:2007vlo, Lopez-Ortega:2012xvr}:
\begin{equation}
\omega^{(\text{dS})}_n r_c = -i\left( \ell + n + 1 - \delta_{n0} \right).
\end{equation}
This expression was generalized to the case with asymptotic behavior \( \omega_q = -2/3 \) in \cite{Konoplya:2025mvj}. 
Furthermore, as the parameters are tuned such that the BH contribution becomes subdominant (i.e., \( M/r_c \ll 1 \)), one observes a smooth transition from BH-dominated quasinormal ringing to a regime where the modes resemble cosmological resonances of the pure de Sitter background. These deformed cosmological modes could potentially serve as probes of the large-scale structure of spacetime in future gravitational wave observations, particularly in scenarios involving DE dynamics.

In the eikonal limit (\( \ell \gg 1 \)), quasinormal modes of BHs are often approximated by the parameters of unstable circular null geodesics~\cite{Cardoso:2008bp}. The real and imaginary parts of the quasinormal frequencies in this regime are given by
\begin{equation}
\omega_{\text{QNM}} = \Omega_c \, \ell - i \left( n + \frac{1}{2} \right) |\lambda|,
\end{equation}
where \( \Omega_c \) is the angular velocity and \( \lambda \) is the Lyapunov exponent associated with the instability of the circular null geodesic. This correspondence accurately describes the Schwarzschild-like branch of modes, even in the presence of quintessence and DM, provided the effective potential retains a dominant single peak structure near the photon sphere. However, it fails for the de Sitter branch of quasinormal modes \cite{Konoplya:2022gjp}, which are governed instead by the asymptotic cosmological structure and cannot be related to null geodesics. Therefore, the geodesic–QNM correspondence holds in this spacetime, but only for the overtones \( n \) of the Schwarzschild-like family.

The quasinormal modes computed using the WKB method for scalar perturbations with $\ell = 0$, $1$, and $2$ are presented in Tables~\ref{QNMtable1}--\ref{QNMtable3}, for various values of $\rho_s$ and $\omega_q$, with $c$ and $r_s$ kept fixed. The choice of Padé approximant is specified by the parameter $m$, which is selected following the prescription in~\cite{Konoplya:2019hlu} to reproduce the Schwarzschild-limit spectrum with optimal accuracy. In the limit of vanishing astrophysical environment, and once $\omega_q=-1,$ we reproduce the quasinormal modes of the Schwarzschild-de Sitter BH \cite{Konoplya:2004uk, Zhidenko:2003wq}.

Although there is no rigorous estimate of the relative error inherent to the WKB method, it is generally accepted that when the difference between successive orders is small, the true error is of the same order as that difference. In Tables~\ref{QNMtable1}--\ref{QNMtable3}, we observe that for $\ell = 1$ and higher, the variation between different WKB orders is much smaller than the effect itself—that is, the deviation of the frequency from its Schwarzschild value. For $\ell = 0$, this is not always the case for certain parameter choices, but even in those instances, the overall trend remains consistent. For example, when $\rho_s$ varies from $0.05$ to $0.5$, the total frequency shift exceeds one percent, while the estimated relative error is about $0.1\%$, i.e., an order of magnitude smaller than the effect. For higher $\ell$, the expected relative error is smaller than the effect by two or more orders of magnitude, making the results sufficiently reliable for our analysis.

We also note that while $\omega_q$ significantly influences the quasinormal frequencies, the density parameter $\rho_s$ has a comparatively minor effect, altering the frequencies by at most one to two percent. A higher density leads to slightly smaller real and imaginary parts of the quasinormal frequency for the Dehnen matter distribution. A similar behavior was observed in~\cite{Konoplya:2021ube} for the Hernquist-type distribution of matter, where increasing the total mass $M$ of the halo (while keeping its radius fixed) resulted in a suppression of both the real and imaginary parts of the frequency.

In addition, we find that the deviation of quasinormal frequencies due to the presence of DM (encoded in the Dehnen profile) becomes more pronounced when combined with stronger quintessence effects (i.e., more negative \( \omega_q \)). This suggests a mild but nontrivial interplay between DM and DE sectors at the perturbative level, hinting that quasinormal modes may carry imprints of their joint distribution in realistic astrophysical settings.

An example of the dependence of the quasinormal frequencies on $r_c$ is presented in Table~\ref{QNMtable4}. We observe that a nonzero negative value of $\omega_q$, which yields a de Sitter-like asymptotic structure, leads to a suppression of both the real part of the oscillation frequency and the damping rate—an effect similar to that seen in various asymptotically de Sitter BH spacetimes~\cite{Zhidenko:2003wq, Cuyubamba:2016cug, Konoplya:2007zx, Dubinsky:2024gwo}.

To validate the above conclusions regarding the expected accuracy of the WKB method, we also compute the quasinormal modes using time-domain integration. However, this approach is significantly more time-consuming, so we restrict ourselves to presenting illustrative results in Fig. \ref{fig:TDL0}. An additional challenge arises in the case of $\ell = 0$, which is not only the most problematic for the WKB method, but also for time-domain integration. In this case, the ringdown stage consists of only a few oscillations, which is insufficient for extracting the frequencies with high accuracy. Consequently, the observed discrepancy between the WKB and time-domain results for $\ell = 0$ should not be interpreted as an indication that the time-domain method is more accurate in this regime. In contrast, for $\ell \geq 1$, the agreement between both methods is excellent (see Fig. \ref{fig:TDL0}).

Once $\omega_q$ becomes negative, the resulting de Sitter-like asymptotic causes quasinormal modes to dominate not only during the ringdown phase but at all times~\cite{Dyatlov:2010hq,Konoplya:2022xid}, including asymptotically late times, where power-law tails would typically dominate in asymptotically flat spacetimes~\cite{Price:1971fb}. This feature enables the extraction of the fundamental mode and even several overtones with unprecedented accuracy~\cite{Dubinsky:2024gwo}. In certain parameter ranges, we also observe the emergence of a de Sitter branch of modes that dominates the late-time signal. An example is shown in Fig. \ref{fig:TDL0}, where the non-oscillatory, exponentially decaying tail corresponds to such a mode. A similar behavior was previously found for the exact de Sitter asymptotic case (i.e., $\omega_q = -1$) in~\cite{Konoplya:2022xid}.

We concentrated here on the Schwarzschild-like branch of modes and do not give detailed data on the de Sitter mode, because it dominates in the late times, when the signal is strongly damped and could not be observed.

\subsection{Greybody Factor}

\begin{figure}
\resizebox{\linewidth}{!}{\includegraphics{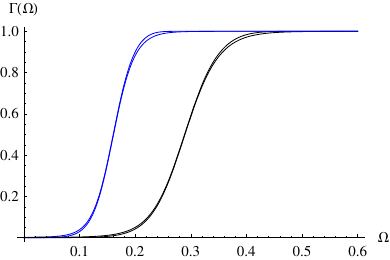}\includegraphics{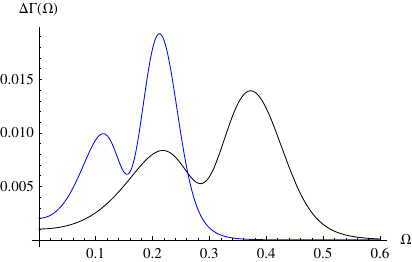}}
\caption{Left panel: Grey-body factors for: $\ell=1$, $M=1$, $r_s =0.1$ (black) and $r_s =0.5$ (blue), $\rho_s =0.03$, $\omega_q =-0.35$, $c=0.01$. Right panel: the difference between grey-body factors obtained via the WKB formula and the correspondence with QNMs.}\label{fig:GBF1}
\end{figure}

\begin{figure}
\resizebox{\linewidth}{!}{\includegraphics{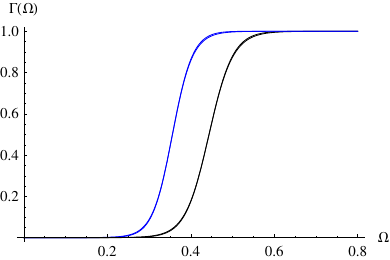}\includegraphics{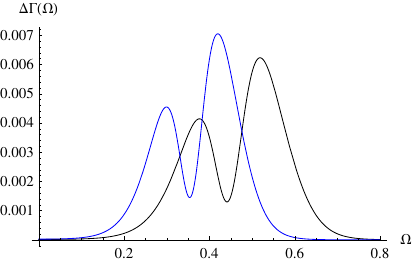}}
\caption{Left panel: Grey-body factors for: $\ell=2$, $M=1$, $\rho_s =0.1$ (black) and $\rho_s =0.5$ (blue), $r_s =0.3$, $\omega_q =-0.35$, $c=0.01$. Right panel: the difference between grey-body factors obtained via the WKB formula and the correspondence with QNMs.}\label{fig:GBF2}
\end{figure}

\begin{figure}
\resizebox{\linewidth}{!}{\includegraphics{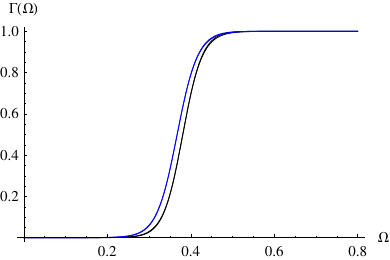}\includegraphics{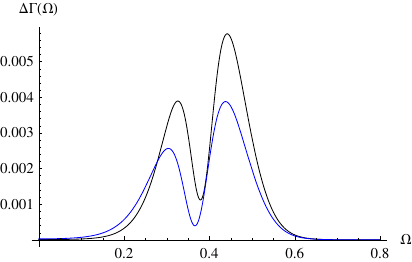}}
\caption{Left panel: Grey-body factors for: $\ell=2$, $M=1$, $\omega_q =-0.35$ (black) and $\omega_q =-1$ (blue), $r_s =0.3$, $\rho_s=0.1$, $c=0.1$. Right panel: the difference between grey-body factors obtained via the WKB formula and the correspondence with QNMs.}\label{fig:GBF3}
\end{figure}

In BH perturbation theory, greybody factors quantify the deviation of the BH’s radiation spectrum from a perfect blackbody, due to the partial reflection of waves by the curvature-induced effective potential outside the event horizon. The greybody factor $\Gamma_\ell(\omega)$ is defined through the scattering problem with the boundary conditions:
\begin{equation}
\Psi(r_*) \sim 
\begin{cases}
e^{-i \omega r_*}, & r_* \to -\infty \quad \text{(ingoing at horizon)} \\
A_{\text{out}} e^{i \omega r_*} + A_{\text{in}} e^{-i \omega r_*}, & r_* \to +\infty \quad \text{(outgoing + ingoing at infinity or de Sitter-like horizon)}
\end{cases}
\end{equation}
The transmission coefficient, or greybody factor, is then given by $\Gamma_\ell(\omega) = |T_\ell(\omega)|^2 = 1 - |R_\ell(\omega)|^2$, where $R_\ell = A_{\text{in}}^{-1}$ and $T_\ell$ is the transmission amplitude through the potential barrier.

Here, we can calculate greybody factors using the 6th order WKB formula of \cite{Konoplya:2003ii}
\[
R = \left(1 + e^{-2i\pi K}\right)^{-1/2},
\]
where $K$ is a function of the frequency $\Omega$ and the effective potential and its derivatives at the maximum until the 12th order. Alternatively, using the correspondence between grey-body factors and quasinormal modes found in~\cite{Konoplya:2024lir}) , the transmission coefficient can be calculated by the following formula:
\[
\Gamma_\ell(\Omega)  = \left(1 + \exp\left[\frac{2\pi\left(\Omega^2 - \text{Re}(\omega_0)^2\right)}{4\, \text{Re}(\omega_0)\, |\text{Im}(\omega_0)|} \right]\right)^{-1} + \mathcal{O}(\ell^{-1}).
\]
Here $\Omega$ is the real frequency associated with the energy of the radiation flow in the scattering problem, and $\omega_0$ is the fundamental quasinormal mode frequency.  Correction terms beyond the eikonal regime include also the first overtone. This correspondence has been recently tested in a number of publications  leading to sufficiently accurate estimates of greybody factors even at relatively low multipole numbers \cite{Malik:2024cgb, Bolokhov:2024otn, Konoplya:2024vuj, Konoplya:2024lch, Pedrotti:2025upg, Hamil:2025cms, Lutfuoglu:2025ljm, Dubinsky:2024vbn, Skvortsova:2024msa}. We will use both methods for calculation of the greybody factors and compare them to test the correspondence. The WKB method is naturally more accurate than the correspondence.

From Figs.~\ref{fig:GBF1} and \ref{fig:GBF2}, we observe that increasing either the halo radius \( r_s \) or the density \( \rho_s \) leads to an enhancement of the greybody factors. This implies that the larger and denser the galactic halo, the more amplified the radiation becomes — provided the radiation does not interact with the halo matter, which is a reasonable assumption in certain DM models. At the same time from fig. \ref{fig:GBF3} we see that the cosmological factor $\omega_q$ does not influence much greybody factors, once the constant $c$ is small enough. 
We also note that the correspondence between greybody factors and quasinormal modes becomes more accurate for higher multipole numbers \( \ell \).


To validate the above results regarding the the greybody factor we will use the semi-analytic limits method to obtain lower bounds for the greybody factor \cite{qn41,qn44} 
\begin{equation}
T\left( \Omega\right) \geq \sec h^{2}\left( \frac{1}{2\Omega}%
\int_{r_{h}}^{+\infty }V_{s}dr_{\ast }\right) ,  \label{is8}
\end{equation}
We will use the perturbative potential of massless scalar given in Eq. (\ref{ff7}) to obtain the bounds of the greybody factor of the  Schwarzschild BHs situated within a Dehnen-type DM halo with quintessence.  Consequently, we obtain the analytical solution of the bounds of the greybody factor as 
\begin{equation}
    T\left( \Omega\right) \geq \sec h^{2}\left[ \frac{1}{\Omega }\left( \frac{32\pi}{3\,r_h}r_s\,\rho_s \left( 2\,r_h\left( 1+\sqrt{1+\frac{r_s}{r_h}} \right)-r_s \sqrt{1+\frac{r_s}{r_h}}  \right) -\frac{l(l+1)%
}{r_{h}}-\frac{M}{r_{h}^{2}}-\frac{c (1+3\omega_{q})}{(2+3\omega_{q})r_{h}^{2+3\omega_{q}
}}\right)
\right]. \label{gf1}
\end{equation}
Figure \ref{gf1} plots transmission coefficients versus $\omega$ for different core density (left) and radius (right).  It is clear that as the parameters $\rho_s$ and $r_s$ grow, so do the transmission coefficients, meaning that more thermal radiation will reach the observer at spatial infinity. Asymptotically, the greybody factor approaches unity, the value of a blackbody.
\begin{figure}[H]
\begin{center}
\includegraphics[scale=0.85]{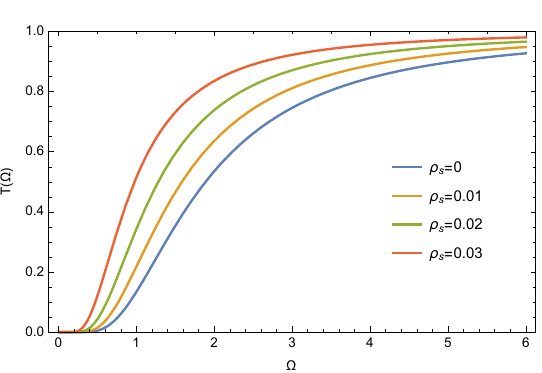}
\includegraphics[scale=0.85]{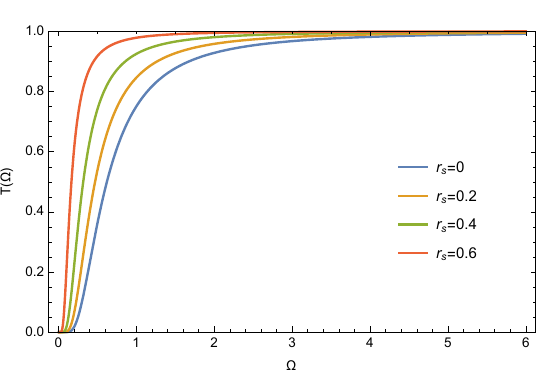} 
\end{center}
\caption{Greybody bounds  of scalar massless field  for different values of  core $\rho_s$ (left) and $r_s$ (right). Here, $M=1$, $w_q=-0.8$ and $c=0.01$. }\label{gf11}
\end{figure}

 \subsection{Sparsity of Hawking radiation}
\begin{figure}[H]
\begin{center}
\includegraphics[scale=0.85]{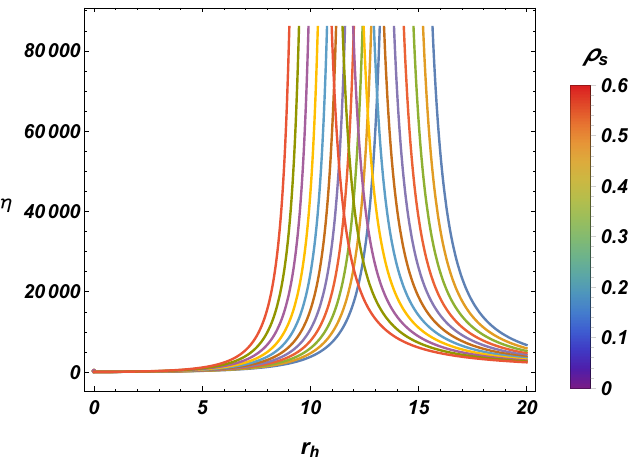}
\includegraphics[scale=0.85]{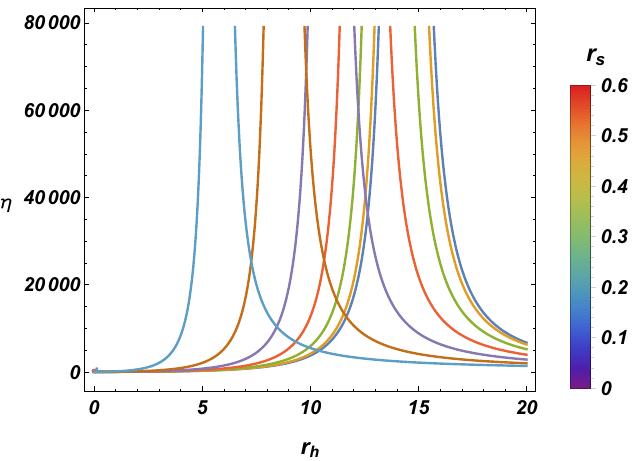} 
\end{center}
\caption{Plot of $\eta $ as a function of $r_h $ for multiple values of the  core density $\rho_s$ (left) and radius  $r_s$ (right). Here, $M=1$, $w_q=-0.8$ and $c=0.01$. }\label{gf12}
\end{figure}

The aim of this part is to determine the sparsity of BHs. In general, a BH behaves similarly to a black body, spewing particles at a temperature relative to surface gravity. Nonetheless, the Hawking radiation flow differs from ordinary blackbody radiation in that it appears very scarce throughout the evaporation phase. The sparity is defined as the average time between the emission of successive quanta on time scales determined by their energy. Sparity is defined as \cite{Page:1976df,Gray:2015pma}
\begin{equation}
\label{defSpars}
    \eta =\frac{\mathcal{C}}{\Tilde{g} }\left(\frac{\lambda_t^2}{\mathcal{A}_{eff}}\right),
\end{equation}
  where the parameter $\mathcal{C}$ is a dimensionless constant, $\Tilde{g}$ stands for the spin degeneracy factor of the emitted quanta, $\lambda_t=2\pi/T$ represents the thermal wavelength, and $\mathcal{A}_{eff}=27 \mathcal{A}_{BH}/4$ is simply the effective area of the BH. For the simple case of a Schwarzschild BH and spin-1 bosons with no emitted mass, this gives $\lambda_t=8\pi r_h^2\,\Longrightarrow\,\eta_{Sch}=64\pi^3/27$. For the sake of comparison, recall that $\eta\ll 1$ for black body radiation.\\ In order to examine the sparsity behavior of Hawking radiation, we consider the Hawking temperature, which is defined as 
  \begin{equation}
    T=\frac{1}{4\pi}\left(\frac{\mathrm{d} \mathcal{F}(r)}{\mathrm{d}r}\right)_{r=r_h}.
\end{equation}
In terms of parameter space, it can be expressed as follows:
\begin{equation}
    T_H=\frac{r_h+r_s+3c\,w_q\,r^{-(1+3w_q)}-16\pi\,r_s^2\,\rho_s\sqrt{1+\frac{r_s}{r_h}}(r_s+2r_h)}{4\pi\,r_h(r_s+r_h)}.
\end{equation}
To expose the physical boundary of the sparsity aspect, we consider the temperature above and the surface area of the horizon, $\mathcal{A}_{BH}=4\pi r_h^2$. Thus, Eq. (\ref{defSpars}) becomes
\begin{equation}
    \eta =\frac{16\pi^2\,r_h^2(r_s+r_h)^2}{27\left(r_h+r_s+3c\,w_q\,r^{-(1+3w_q)}-16\pi\,r_s^2\,\rho_s\sqrt{1+\frac{r_s}{r_h}}(r_s+2r_h)   \right)^2}. \label{sp32} 
\end{equation}
To fully understand the sparsity behaviour of our BH solution, we must examine the graphical representation of the $\eta(r_h)$ function in Fig. \ref{gf12}. This output reveals two unique scenarios for sparsity behaviour based on the suitable sign of the $\eta(r_h)$ function. When $r_h$ is sufficiently large, the corresponding sparsity exceeds the usual $\eta_{Sch}$. This suggests that the radiation being emitted is denser than Hawking radiation at this stage of the evaporation process. As $r_h$ grows, $\eta$ declines monotonically and approaches zero. At this rate, the behaviour begins to closely resemble that of black-body radiation.

\section{Conclusion} \label{sec:conclusion}
In this study, we comprehensively investigated the Schwarzschild BHs embedded
in a Dehnen-type DM halo with QF. Our analysis provided significant insights into how the DM, and DE  collectively influence the physical properties of this extended BH geometry.\\ We first derived the
 metric function and discussed its geometric properties, including curvature invariants and horizon structure.  We found that the solution of the BH
 singularity at $r=0$ is a necessary singularity that cannot
 be removed by coordinate translation. The BH defined by metric (\ref{bb1}) has two 
 horizons: the event and the cosmological ones. Both horizons 
 are mostly reliant on the parameters of the dark sector. It 
 was discovered that all dark sector parameters including the 
 quintessence parameter c cause the event horizon to expand while shrinking the cosmological 
 horizon.\\ Our investigation of geodesic motion, including ISCO, photon sphere radii, and BH 
 shadow features, revealed significant influences of both the core density parameter 
 $\rho_s$ and the quintessence parameter $\omega_q$ on the paths 
 of test particles. For time-like geodesics motion, we derived the effective potential given by Eq. 
 (\ref{effp1}), which incorporates the combined effects of DM and DE. Our analysis, visualized in 
 Fig. \ref{fig:typesoforbits}, demonstrated  the trajectories of the radial particle in detail, focusing on various types of orbits corresponding to
different energy values $E$. Next, we  examined the behavior of timelike particles in circular orbits around a BH. The ISCO is 
 also investigated and solved numerically. It is found  that,  the ISCO increases with both core density parameter and the QF 
 parameter. We then extended our analysis to null geodesic motion of photons around the BH. 
 Forevermore, we numerically obtained both the photon sphere and the shadow radii. It is found that both radii increase with core density and QF parameter.\\
A significant component of our 
 research was the examination of the perturbative potential for scalar fields in this BH background. Figure \ref{potScalar} illustrates the 
 surprising sensitivity of the scalar perturbative potential, 
 generated by Equation (\ref{ff7}), to core density and radius parameters.  These discoveries are particularly 
 essential for understanding the stability of the BH under various sorts of perturbations and 
  characterizing its quasinormal mode spectrum. Next, QNMs frequencies are computed Using both
the WKB approximation and time-domain integration. We have analyzed how DM and DE affect the oscillation frequencies and damping rates of these perturbations. We found that while $\omega_q$ has a large influence on quasinormal frequencies, the density parameter $\rho_s$ has a comparably little effect, modifying the frequencies by no more than one to two percent. Higher core density results in slightly smaller real and imaginary components of the quasinormal frequency for the Dehnen matter distribution. We also found that  the deviation of quasinormal frequencies due to the presence of DM (encoded in the
Dehnen profile) becomes more pronounced when combined with stronger quintessence effects. \\ We then computed the quasinormal modes using time-domain integration.  We found a mismatch between the WKB and time-domain results for $\ell = 0$. However, for $\ell \geq 1$, the agreement between both approaches is high (see Fig. \ref{fig:TDL0}). \\ We estimated the greybody factors of these BHs using two methods: high-order JWKB and semi-analytic limitations. Both methods demonstrate great agreement.  We found that increasing the halo radius $r_s$ or density $\rho_s$ enhances the greybody factors. This indicates that the larger and denser the galactic halo, the more magnified the radiation – assuming the radiation does not interact with the halo matter, which is a plausible assumption in some DM models. \\ We then investigated  the sparsity of Hawking radiation. Two distinct possibilities for sparsity behaviour are identified based on the appropriate sign of the $\eta(r_h)$ function. When $r_h$ is large enough, the resulting sparsity exceeds the normal $\eta_{Sch}$. This implies that the radiation being emitted is denser than Hawking radiation at this point in the evaporation process. As $r_h$ increases, $\eta$ decreases steadily and approaches zero.\\Finally, Our study suggests numerous intriguing future research topics. For example, analysing the thermodynamic parameters of Schwarzschild BH embedded
in a Dehnen-type dark matter halo with quintessence. Moreover, extending our research to include rotating BHs would provide a more full understanding of astrophysical BHs, which often exhibit high angular momentum \cite{izz31}.

\section*{Acknowledgments}

 B. C. L. is grateful to Excellence project PřF UHK 2205/2025-2026 for the financial support.

\end{document}